\documentclass[11pt,a4paper]{article}
\pdfoutput=1
\usepackage{jheppub}

\usepackage{graphicx} 
\usepackage{amsmath,amsfonts,amssymb}

\newcommand\fft[2]{\frac{#1}{#2}}

\newcommand\nn{\nonumber}
\DeclareMathOperator\Tr{Tr}
\DeclareMathOperator\PE{PE}

\preprint{LCTP-24-12}

\title{Analytic continuation and the giant graviton expansion}

\author[a]{Nizar Ezroura,}
\author[a]{James T. Liu}
\author[a,b]{and Neville Joshua Rajappa}

\emailAdd{nezroura@umich.edu}
\emailAdd{jimliu@umich.edu}
\emailAdd{nevjoraj@umich.edu}

\affiliation{${}^a$ Leinweber Center for Theoretical Physics, Randall Laboratory of Physics\\The University of Michigan, Ann Arbor, MI 48109-1040, USA }

\affiliation{${}^b$ Department of Physics and Astronomy, Stony Brook University,\\Stony Brook, NY, 11794-3800, USA }

\abstract{There is now strong evidence that the superconformal index of holographic gauge theories admits a giant graviton expansion capturing the worldvolume dynamics of spherical D3-branes moving in internal space.  While the giant graviton indices may be written as suitable contour integrals, care must be taken in their evaluation, as there can be an ambiguity in choice of integration contour or, equivalently, ``pole selection''.  We resolve this ambiguity by using Murthy's recent expansion of the superconformal index to provide a rigorous underpinning of the evaluation of the supergravity giant graviton indices.  In doing so, we directly relate Murthy's giant graviton expansion to the supergravity D3-brane expansion.}

\begin{document}
\maketitle
\flushbottom

\section{Introduction and Summary}

While the AdS/CFT conjecture is generally thought of as a large-$N$ duality, in its strong form it posits an complete equivalence between the boundary field theory and the dual string theory in AdS.  In this case, AdS/CFT holds not just in the large-$N$ limit, but also at finite $N$.  This has led to numerous investigations of precision holography focused on understanding the subleading in $N$ corrections on both sides of the duality.

The study of finite-$N$ corrections has often involved a large-$N$ expansion where the subleading terms have the form of a $1/N$ expansion with the possible addition of $\log N$ terms.  However, recent investigations of the superconformal index have shown that many results can be obtained directly even for $N$ of $\mathcal O(1)$ \cite{Agarwal:2020zwm,Murthy:2020rbd}.  One of the reasons this comparison is even possible is that the index counts states that are protected by supersymmetry, thus keeping the duality under control even in the strong coupling limit.

The superconformal index is a Witten index refined by the addition of fugacities associated with conserved charges commuting with the Hamiltonian.  As shown in \cite{Romelsberger:2005eg,Kinney:2005ej}, for a $U(N)$ gauge theory that admits a weakly coupled limit, we can obtain the index by starting with the single letter index, $f(\mathfrak q)$, combined with the adjoint character, $\chi_{\mathrm{adj}}(U)=\Tr(U)\Tr(U^{-1})$.  The full index, $Z_N(\mathfrak q)$, is then obtained by Plethystically exponentiating $f(\mathfrak q)\chi_{\mathrm{adj}}(U)$ and projecting onto gauge singlets
\begin{equation}
    Z_N(\mathfrak q)=\int_{U(N)}dU\exp\left(\sum_{k\ge1}\fft{f(\mathfrak q^k)}k\Tr(U^k)\Tr(U^{-k})\right).
\label{eq:matrixindex}
\end{equation}
Here the set of fugacities are generically denoted by $\mathfrak q=\{q_1,q_2,\ldots\}$.  In the infinite-$N$ limit, the projection over gauge singlets does not involve any trace relations, and one obtains a relatively simple expression \cite{Kinney:2005ej}
\begin{equation}
    Z_\infty(\mathfrak q)=\prod_{k\ge1}\fft1{1-f(\mathfrak q^k)}.
\end{equation}
This result is easily understood from a combinatorial point of view in combining single letters into the full multi-letter index.

At finite $N$, however, the situation is more involved.  Taken as a polynomial expansion in $\mathfrak q$, at low orders, $k<N$, the finite-$N$ index $Z_N(\mathfrak q)$ agrees with $Z_\infty(\mathfrak q)$.  However, beyond this order, trace relations enter the projection onto singlets, and the expressions will become distinct.  Assuming the single letter index $f(\mathfrak q)$ has an expansion starting with $\mathcal O(\mathfrak q)$, we thus have
\begin{equation}
    \fft{Z_N(\mathfrak q)}{Z_\infty(\mathfrak q)}=1+\mathcal O(\mathfrak q^N).
\end{equation}
In many cases, this can be made more precise.  In particular, we write
\begin{equation}
     \fft{Z_N(q)}{Z_\infty(q)}=1+\sum_{m\ge1}q^{mN} \hat Z_{m}(q),
\label{eq:qGGE}
\end{equation}
for a single fugacity $q$, with obvious generalizations for multiple fugacities, $\mathfrak q$.  This remarkable expansion is known as the giant graviton expansion \cite{Bourdier:2015wda,Bourdier:2015sga,Arai:2019aou,Arai:2020uwd,Imamura:2021ytr,Gaiotto:2021xce,Murthy:2022ien,Lee:2022vig}, and provides important insights into the structure of holographic field theories.

On the field theory side, one can understand the deviation away from $Z_\infty(\mathfrak q)$ as arising from determinant operators which only start contributing at $\mathcal O(\mathfrak q^N)$.  However, the main insight and origin of the name `giant graviton expansion' comes from the gravity side of the duality.  While the full setup is more general, here we focus on $\mathcal N=4$ super-Yang-Mills which is dual to IIB string theory on AdS$_5\times S^5$.  In this case, Kaluza-Klein spectroscopy allows us to identify the Kaluza-Klein graviton states of IIB supergravity on $S^5$ with the states counted by $Z_\infty(q)$.  However, at sufficiently large angular momentum on $S^5$, the modes expand into a spherical D3-brane wrapping $S^3$ in $S^5$ and stabilized by angular momentum; these are the giant gravitons \cite{McGreevy:2000cw}.  The index $\hat Z_m(q)$ in (\ref{eq:qGGE}) is then the index for the worldvolume theory of a stack of $m$ D3-branes wrapped on $S^3$ carrying angular momentum in $S^5$ \cite{Arai:2019aou,Arai:2020uwd,Imamura:2021ytr}.

The beauty of the giant graviton expansion is that both $Z_N(q)$ and $\hat Z_m(q)$ are $\mathcal N=4$ SYM indices (at least for what we focus on).  Thus, in some sense, they are the same object.  This notion that ``$Z_N=\hat Z_m$'' has to be refined, of course.  For one thing, in addition to difference in rank ($N$ vs $m$), the BPS fluctuations of the wrapped D3-branes may carry different charges than those of the original stack of $N$ parallel branes contributing to $Z_N(q)$.  Hence the fugacity $\hat q$ in $\hat Z_m(\hat q)$ may not be the same as $q$ in $Z_N(q)$.  Nevertheless, they are related, as the entire system transforms under a single set of symmetries.  The other important difference is that we ought to consider motion of the giant gravitons in three orthogonal rotation planes corresponding to $U(1)^3$ contained in the R-symmetry group $SU(4)_R$ of $\mathcal N=4$ SYM.  Thus the giant graviton index can be more precisely written as $\hat Z_{(m_1,m_2,m_3)}(\mathfrak q)$, where $m_i$ corresponds to the number of D3-branes moving in the $i$-th rotation plane.  In this case, the giant graviton expansion takes the form
\begin{equation}
    \fft{Z_N(\mathfrak q)}{Z_\infty(\mathfrak q)}=\sum_{m_i\ge0}(q_1^{m_1}q_2^{m_2}q_3^{m_3})^N\hat Z_{(\vec m)}(\mathfrak q).
\end{equation}
Here we have singled out the three fugacities $q_1$, $q_2$ and $q_3$ corresponding to the $U(1)^3$ charges.  The factor $(q_1^{m_1}q_2^{m_2}q_3^{m_3})^N$ is the contribution to the index from `classical' motion of the D3-branes along the rotation planes, while $\hat Z_{(\vec m)}(\mathfrak q)$ corresponds to their worldsheet fluctuations and takes the form of an index for a three-node quiver with ranks $m_i$.  While this is the general case, not all indices need make use of all three rotation planes.

This simplest case to consider is the 1/2 BPS index, which can be worked out explicitly.  For the 1/2 BPS index, with single letter index $f(q)=q$, we have 
\begin{equation}
    Z_N(q)=\fft1{(q;q)_N},
\label{eq:1/2BPSqPoch}
\end{equation}
where the $q$-Pochhammer symbol is defined by 
\begin{equation}
    (z;q)_n=\prod_{k=0}^{n-1}(1-zq^k).
\label{eq:qPoch}
\end{equation}
Here there is a single fugacity, and the 1/2 BPS giant graviton expansion makes use of a single rotation plane to count $\Delta=J$ states so it can be written as
\begin{equation}
    \frac{Z_N(q)}{Z_\infty(q)}=1+\sum_{m>0}q^{mN}\hat Z_m(q).
\label{eq:1/2BPSGGE}
\end{equation}
As a result, the giant graviton index
\begin{equation}
    \hat Z_m(q)=\fft1{(q^{-1};q^{-1})_m}=Z_m(q^{-1}),
\end{equation}
is that of a single node quiver with gauge group $U(m)$ and opposite R-charge, $-J$ \cite{Gaiotto:2021xce}.

While the relation $\hat Z_m(q)=Z_m(q^{-1})$ is manifest in terms of the $q$-Pochhammer symbols, there is nevertheless an important subtlety that arises when Plethystically exponentiating the single letter index according to (\ref{eq:matrixindex}).  For $f(q)=q$, we can write the explicit expression
\begin{align}
    Z_N(q)=\fft1{N!}\int\prod_i\fft{dz_i}{2\pi iz_i}\fft{\prod_{i\ne j}1-z_{ij}}{\prod_{i,j}1-qz_{ij}},
\label{eq:1/2BPSintegral}
\end{align}
where $z_{ij}=z_i/z_j$, and where we have assumed $|q|<1$ for convergence of the index.  The contours are taken on the unit circle, and the multi-dimensional contour integral essentially picks up contributions only from poles contained within the contour.  Careful evaluation of the integral then gives the result (\ref{eq:1/2BPSqPoch}).

So far, all is well for evaluating $Z_N(q)$.  However, an issue arises when considering the giant graviton contribution $\hat Z_m(q)$ arising from the Plethystic exponential of the single letter index $f_v(q)=q^{-1}$.  Assuming still that $|q|<1$, in this case the replacement $q\to q^{-1}$ interchanges poles inside and outside the unit contour.  In particular, as we will show, making the substitution $q\to q^{-1}$ in the integral (\ref{eq:1/2BPSintegral}) and evaluating with the same unit contour as below gives
\begin{equation}
    Z_N(q^{-1})=\fft{q^{N(N-1)/2}}{(q^{-1};q^{-1})_N},
\label{eq:1/2BPScontourint}
\end{equation}
which is not the same as simply substituting $q\to q^{-1}$ in (\ref{eq:1/2BPSqPoch}).  The point here is that the result of a contour integral of the form (\ref{eq:1/2BPSintegral}) with movable poles will certainly depend on which set of poles are included within the contour.

This issue of which poles to include in the contour is a known issue facing the giant graviton expansion.  One may assume that the natural choice would be to take all integration variables $z_i$ on the unit circle and to only pick up contributions from the poles within the unit contour as one ordinarily does in contour integration.  However, this would give (\ref{eq:1/2BPScontourint}), which does \textit{not} yield a valid giant graviton expansion of the form (\ref{eq:1/2BPSGGE}).  Instead, the appropriate choice here is to simply take the original 1/2 BPS index, (\ref{eq:1/2BPSqPoch}), and make the substitution $q\to q^{-1}$.  This choice can be viewed in multiple ways.  One is to take (\ref{eq:1/2BPSqPoch}) as the \textit{full} index for all fugacities $q$, whether inside or outside the unit circle.  This can be thought of as an `analytic continuation' definition of the index, and is the approach taken in \cite{Gaiotto:2021xce,Lee:2022vig,Imamura:2022aua}.  Another is to keep the contour integral definition of the index but suitably deform the contour so that no poles cross the contour when taking $q\to q^{-1}$.  This approach is essentially that taken in \cite{Arai:2020qaj,Imamura:2021ytr}, where it was denoted the ``pole selection rule''.

Since ``pole selection'' is basically analytical continuation in disguise, one may simply declare that the giant graviton expansion ought to be done by analytic continuation.  However, while this is straightforward for the 1/2 BPS index, in other cases exact expressions for the index $Z_N(\mathfrak q)$ may not be known.  Instead, one may have to work with series expansions of the index, and analytic continuation of a series outside of its radius of convergence is not necessarily straightforward especially if the series is only incompletely specified.  In this case, there may appear to be little choice but to fall back on pole selection and its related issues.  While this may have been justified in a somewhat \textit{ad hoc} manner, the goal of this paper is to demonstrate that the giant graviton expansion can be made rigorous through the use of Murthy's expansion \cite{Murthy:2022ien} of the unitary matrix model expression (\ref{eq:matrixindex}).

As demonstrated in \cite{Murthy:2022ien}, a clever re-expression of the matrix integral in terms of a system of free fermions yields a rigorous expansion of the form
\begin{equation}
    \fft{Z_N(\mathfrak q)}{Z_\infty(\mathfrak q)}=1+\sum_{m\ge1}G_N^{(m)}(\mathfrak q),
\label{eq:MurthyGGE}
\end{equation}
where $G_N^{(m)}(\mathfrak q)$ takes the form of a fermionic determinant and scales as $\mathcal O(\mathfrak q^{mN})$.  This expansion has also been denoted a giant graviton expansion and indeed has the same form as (\ref{eq:qGGE}).  However, as demonstrated explicitly in \cite{Liu:2022olj}, the two expansions are \textit{not} the same.  Comparison of (\ref{eq:MurthyGGE}) with (\ref{eq:qGGE}) suggests that Murthy's $G_N^{(m)}(q)$ should be identified with wrapped D3-brane index $q^{mN}\hat Z_m(q)$, at least in the case of a single fugacity.  Yet this turns out not to be the case, not even for the straightforward 1/2 BPS index.  Nevertheless, as demonstrated in \cite{Eniceicu:2023uvd}, there is a way to relate the two expansions such that $\hat Z_m(q)$ receives contributions from all kernels $G_N^{(m')}(q)$ with $1\le m'\le m$.

While the relation between $G_N^{(m)}(q)$ and the D3-brane index $\hat Z_m(q)$ is not at all straightforward, the advantage of working out this relation is that $G_N^{(m)}(q)$ can be rigorously defined without pole selection ambiguity.  In this way, Murthy's giant graviton expansion provides a rigorous underlying definition of the D3-brane giant graviton expansion, and moreover can serve as an explicit computational tool for further investigations of the index.  This connection between expansions is also notable in that the unitary matrix integral, (\ref{eq:matrixindex}), taken as a starting point of Murthy's expansion makes no reference to giant gravitons nor to D3-branes.  And yet such objects naturally arise once the index is reorganized in terms of $\mathcal O(q^N)$ expressions.  In a way, this highlights the connection between determinant operators in the $U(N)$ gauge theory and their giant graviton counterparts.

In this paper we connect the prevailing two giant graviton expansions in the literature, thus providing a unified description of the giant graviton index and a more rigorous approach without having to appeal to any pole selection rules.  In particular, what we demonstrate are the following:
\begin{enumerate}
    \item We refine the work of \cite{Murthy:2022ien,Liu:2022olj,Eniceicu:2023uvd} to obtain an explicit matrix integral expression for $G_N^{(m)}(\mathfrak q)$ that is well suited to further study.
    \item After a brief summary of the 1/2 BPS index considered in \cite{Eniceicu:2023uvd}, we turn to the 1/8 BPS (Schur) index and show that the double poles hinted at in \cite{Eniceicu:2023uvd} may be avoided in the flavored Schur index, $Z_N(x,y)$.  With this in mind, we are then able to partially evaluate Murthy's $G_N^{(m)}(x,y)$ and split them up into individual contributions towards the D3-brane indices $\hat Z_{(m_1,m_2)}(x,y)$.
    \item We also consider the expansion of the 1/16 BPS index, but leave additional details to future work.  Nevertheless, we demonstrate how the previous results may be recovered in the Schur limit of the 1/16 BPS index.
\end{enumerate}

The plan for the rest of this paper is as follows.  In Section~\ref{sec:murthy} we summarize Murthy's giant graviton expansion and provide a convenient expression for $G_N^{(m)}(\mathfrak q)$ that will be used subsequently.  In Section~\ref{sec:1/2}, we use the giant graviton expansion of the 1/2 BPS index to motivate the distinction between analytic continuation and contour integration when evaluating wrapped D3-brane indices with fugacities outside the unit circle.  We then turn to the Schur index in Section~\ref{sec:Schur} and conclude with the 1/16 BPS index in Section~\ref{sec:1/16}.

\section{Murthy's unitary matrix integral expansion}
\label{sec:murthy}

In the standard approach to the index, where the theory admits a weakly coupled limit, one starts with a single letter index, $f(\mathfrak q)$, and then obtains the full index by its Plethystic exponential, as in (\ref{eq:matrixindex}).  While this integral in itself does not obviously admit an expansion in terms of wrapped D3-branes, in \cite{Murthy:2022ien}, Murthy nevertheless demonstrated how it can be manipulated into the form of a giant graviton type expansion, (\ref{eq:MurthyGGE}).  In particular, the terms in the expansion, $G_N^{(m)}(\mathfrak q)$, are computed from the single letter index, $f(\mathfrak q)$, and scales as $G_N^{(m)}(\mathfrak q)\sim\mathfrak q^{\alpha m(N+m)}$, where $\alpha$ is the leading power in the $f(\mathfrak q)\sim\mathfrak q^\alpha$ expansion \cite{Murthy:2022ien,Liu:2022olj}.

An explicit expression for computing the one-giant contribution, $G_N^{(1)}(\mathfrak q)$, was originally provided in \cite{Murthy:2022ien}, and this was extended to multiple giants in \cite{Liu:2022olj}.  In particular, we can write
\begin{equation}
    G_N^{(m)}(\mathfrak q)=\fft{(-1)^m}{m!}\int\prod_i\left[\fft{dz_i}{2\pi iz_i^2}\fft{dw_i}{2\pi i}\fft{(w_i/z_i)^N}{1-w_i/z_i}\right]\det\left(\fft1{1-w_j/z_i}\right)\exp\left(-\sum_{k=1}^\infty\fft{\gamma_k}k\alpha_k\beta_k\right),
\end{equation}
where
\begin{equation}
    \alpha_k=\sum_iz_i^k-w_i^k,\qquad\beta_k=\sum_iz_i^{-k}-w_i^{-k},\qquad\gamma_k=\hat f(\mathfrak q^k)\equiv\fft{f(\mathfrak q^k)}{1-f(\mathfrak q^k)}.
\end{equation}
As highlighted in \cite{Eniceicu:2023uvd}, this can be rewritten as
\begin{equation}
    G_N^{(m)}(\mathfrak q)=\fft{(-1)^m}{(m!)^2}\int\prod_i\left[\fft{dz_i}{2\pi iz_i^2}\fft{dw_i}{2\pi i}\left(\fft{w_i}{z_i}\right)^N\right]\det\left(\fft1{1-w_j/z_i}\right)^2\exp\left(-\sum_{k=1}^\infty\fft{\gamma_k}k\alpha_k\beta_k\right).
\label{eq:EnGG}
\end{equation}
The advantage of this expression is that it is explicitly permutation invariant on the $\{w_i\}$ and the $\{z_i\}$ variables.

We note that the exponential in (\ref{eq:EnGG}) is just a Plethystic exponential of $\hat f(\mathfrak q)$ dressed by the product
\begin{equation}
    -\alpha_1\beta_1=\sum_{i,j}\left(\fft{z_i}{w_j}+\fft{w_i}{z_j}-\fft{z_i}{z_j}-\fft{w_i}{w_j}\right).
\end{equation}
This allows us to write
\begin{equation}
    \exp\left(-\sum_{k=1}^\infty\fft{\gamma_k}k\alpha_k\beta_k\right)=\prod_{i,j}\PE\left(\hat f(\mathfrak q)\fft{z_i}{w_j}+\hat f(\mathfrak q)\fft{w_i}{z_j}-\hat f(\mathfrak q)\fft{z_i}{z_j}-\hat f(\mathfrak q)\fft{w_i}{w_j}\right).
\label{eq:PEexp}
\end{equation}
Moreover, the determinant in (\ref{eq:EnGG}) can be rewritten in a product form using the Cauchy determinant identity
\begin{equation}
    \det\left(\fft1{1-w_j/z_i}\right)^2=\prod_i\left(\fft{w_i}{z_i}\right)^{m-1}\fft{\prod_{i\ne j}(1-\fft{z_i}{z_j})(1-\fft{w_i}{w_j})}{\prod_{i,j}(1-\fft{w_j}{z_i})^2}.
\label{eq:detsq}
\end{equation}
As a result, we arrive at \cite{Eniceicu:2023uvd}
\begin{align}
    G_N^{(m)}(\mathfrak q)&=\fft{(-1)^m}{(m!)^2}\int\prod_i\left[\fft{dz_i}{2\pi iz_i}\fft{dw_i}{2\pi iw_i}\left(\fft{w_i}{z_i}\right)^{N+m}\right]\fft{\prod_{i\ne j}(1-\fft{z_i}{z_j})(1-\fft{w_i}{w_j})}{\prod_{i,j}(1-\fft{w_j}{z_i})^2}\nn\\
    &\kern10em\times\prod_{i,j}\PE\left(\hat f(\mathfrak q)\fft{z_i}{w_j}+\hat f(\mathfrak q)\fft{w_i}{z_j}-\hat f(\mathfrak q)\fft{z_i}{z_j}-\hat f(\mathfrak q)\fft{w_i}{w_j}\right).
\label{eq:EnGG2}
\end{align}
This expression will be the basis of further manipulation below when we consider the various 1/2, 1/8 and 1/16 BPS indices.

In order to proceed with the evaluation of (\ref{eq:EnGG2}), we need to specify $\hat f(\mathfrak q)$, which depends on the single letter index $f(\mathfrak q)$ according to $\hat f(\mathfrak q)=f(\mathfrak q)/(1-f(\mathfrak q))$.  In general, we expand $\hat f(\mathfrak q)$ into a sum of monomials in the schematic form $f(\mathfrak q^k)=\mathfrak q+\mathfrak q^2+\mathfrak q^3+\cdots$, and use the fact that
\begin{equation}
    \PE(x)=\fft1{1-x},
\label{eq:PE(x)}
\end{equation}
to rewrite the integrand in (\ref{eq:EnGG2}) in a form that is amenable to direct computation.

\subsection{Giant graviton indices from D3-branes}

The most general case we consider is that of the 1/16 BPS index which can be obtained from the single letter index
\begin{equation}
    f(p,q;y_i)=1-\frac{(1-y_1)(1-y_2)(1-y_3)}{(1-p)(1-q)};\qquad pq=y_1y_2y_3.
\label{eq:f1/16}
\end{equation}
Here we have taken the three R-charge fugacities to be $y_i$, so the 1/16 BPS giant graviton expansion takes the form
\begin{equation}
    \fft{Z_N(p,q;y_i)}{Z_\infty(p,q;y_i)}=\sum_{m_i\ge0}(y_1^{m_1}y_2^{m_2}y_3^{m_3})^N\hat Z_{(m_1,m_2,m_3)}(p,q;y_i),
\end{equation}
where $\hat Z_{(m_1,m_2,m_3)}(p,q;y_i)$ is the index of the three-node quiver built using the vector multiplet and hypermultiplet single letter indices \cite{Imamura:2021ytr,Gaiotto:2021xce,Lee:2022vig}
\begin{equation}
    \begin{aligned}
    f_v^{(a)}(p,q;y_a,y_b,y_c)&=1-\fft{(1-1/y_a)(1-p)(1-q)}{(1-y_b)(1-y_c)},\\
    f_h^{(a,b)}(p,q;y_a,y_b,y_c)&=\left(\fft1{y_ay_b}\right)^{1/2}\fft{(1-p)(1-q)}{1-y_c},
    \end{aligned}
    \kern4em(a\ne b\ne c).
\label{eq:1/16singleletter}
\end{equation}
Here $a$, $b$ and $c$ label the nodes of the quiver.  Note, in particular, the mapping of the fugacities for the vectors
\begin{equation}
    f_v^{(a)}(p,q;y_a,y_b,y_c)=f(y_b,y_c;y_a^{-1},p,q),\qquad(a\ne b\ne c),
\label{eq:fvamap}
\end{equation}
where $f(\cdots)$ on the right-hand side is the original single letter index, (\ref{eq:f1/16}).

We can also consider the 1/8 BPS, or Schur, index, with can be obtained from the 1/16 BPS index by taking the fugacities
\begin{equation}
    \{p,q;y_1,y_2,y_3\}\to\{xy,q;x,y,q\}.
\end{equation}
In this case, $q$ drops out, and we end up with the corresponding single letter index
\begin{equation}
    f(x,y)=1-\frac{(1-x)(1-y)}{(1-xy)}.
\end{equation}
In this limit, the third rotation plane (corresponding to fugacity $y_3$) drops out, and only the two remaining rotation planes are relevant.  The giant graviton expansion of the Schur index thus takes the form
\begin{equation}
    \fft{Z_N(x,y)}{Z_\infty(x,y)}=\sum_{m_i\ge0}(x^{m_1}y^{m_2})^N\hat Z_{(m_1,m_2)}(x,y).
\end{equation}
Here, the giant graviton index is that of a two-node quiver with single letter indices
\begin{equation}
    f_v^{(1)}(x,y)=f(x^{-1},xy),\qquad f_v^{(2)}(x,y)=f(xy,y^{-1}),\qquad f_h^{(1,2)}(x,y)=\left(\fft1{xy}\right)^{1/2}(1-xy),
\end{equation}
and takes the form
\begin{equation}
    \hat Z_{(m_1,m_2)}(x,y)=Z_{m_1}(x^{-1},xy)(xy)^{m_1m_2}Z_{m_2}(xy,y^{-1}),
\end{equation}  
where $Z_{m_i}(\cdot,\cdot)$ is just $Z_N(x,y)$ with appropriate identification of rank and fugacities.

Finally, the 1/2 BPS index can be considered the limiting case, $y=0$, of the Schur index.  Renaming $x\to q$ then gives the single letter index $f(q)=q$ and the giant graviton expansion (\ref{eq:1/2BPSGGE}) given in the introduction.  The wrapped D3-brane giant graviton index, $\hat Z_m(q)$, is that of a single node quiver with vector multiplet contribution $f_v(q)=f(q^{-1})$.

From the holographic point of view, the giant graviton indices, $\hat Z_{(m_1,m_2,m_3)}(p,q;y_i)$, $\hat Z_{(m_1,m_2)}(x,y)$ and $\hat Z_m(q)$, are those of the wrapped D3-brane worldvolume theory.  How this is related to Murthy's expansion, (\ref{eq:EnGG2}), which has no obvious D3-brane interpretation, is what we address below.

\section{The 1/2 BPS index}
\label{sec:1/2}

While the 1/2 BPS index is generally well understood, and was investigated in detail in \cite{Eniceicu:2023uvd}, it is instructive to revisit this case before turning to the 1/8 and 1/16 BPS indices.  Recall that the 1/2 BPS index is a function of a single fugacity, $q$, and can be obtained by Plethystically exponentiating the single letter index, $f(q)=q$, according to (\ref{eq:matrixindex}).  Explicitly, we have
\begin{equation}
    Z_N(q)=\fft1{N!}\fft1{(1-q)^N}\int\prod_i\fft{dz_i}{2\pi iz_i}\prod_{i\ne j}\fft{1-z_{ij}}{1-qz_{ij}}.
\label{eq:PE1/2}
\end{equation}
Here the integrals are taken over the unit circle, and we have assumed $|q|<1$.  Evaluating the contour integrals with $|q|<1$ gives the familiar 1/2 BPS $U(N)$ index
\begin{equation}
    Z_N(q)=\fft1{(q;q)_N},
\label{eq:ZN(q)}
\end{equation}
where the $q$-Pochhammer symbol was defined in (\ref{eq:qPoch}).

\subsection{The D3-brane giant graviton expansion}

The infinite-$N$ limit of the index is then $Z_\infty(q)=1/(q;q)_\infty$, and we can obtain the giant graviton expansion
\begin{equation}
    \fft{Z_N}{Z_\infty}=(q^{N+1};q)_\infty=\sum_{m=0}^\infty(-)^mq^{mN}\fft{q^{m(m+1)/2}}{(q;q)_m}.
\end{equation}
This is well-defined for $|q|<1$, and takes the form of a giant graviton expansion
\begin{equation}
    \fft{Z_N(q)}{Z_\infty(q)}=1+\sum_{m=1}^\infty q^{mN}\hat Z_m(q),
\end{equation}
where
\begin{equation}
    \hat Z_m(q)=(-1)^m\fft{q^{m(m+1)/2}}{(q;q)_m}=\fft1{(q^{-1};q^{-1})_m}.
\label{eq:hatZ_m(q)}
\end{equation}
In particular, we observe that \cite{Gaiotto:2021xce}
\begin{equation}
    \hat Z_m(q)=Z_m(q^{-1}).
\label{eq:GLGGE}
\end{equation}
Here, $\hat Z_m(q)$ is identified as the $m$ giant graviton index, while $Z_m(q^{-1})$ is the $U(m)$ super-Yang-Mills index, (\ref{eq:ZN(q)}). with fugacity $q$ replaced by $q^{-1}$.

At this point, it is important to make note of the pole selection procedure in the contour integral for the index, (\ref{eq:PE1/2}).  Since we integrate $\{z_i\}$ on the unit circle, we pick up all residues inside the unit circle.  This is actually somewhat subtle for a multidimensional contour integral, as there can be issues with multivariate residues.  Nevertheless, for the present case, we formally take the result of the contour integration to be the $q$-Pochhammer expression, (\ref{eq:ZN(q)}).  Now consider the replacement $q\to q^{-1}$ in the contour integral.  Here we still assume $|q|<1$, so that $|q^{-1}|>1$.  In this case, the integral expression becomes
\begin{align}
    Z_N(q^{-1})&=\fft1{N!}\fft1{(1-q^{-1})^N}\int\prod_i\fft{dz_i}{2\pi iz_i}\prod_{i\ne j}\fft{1-z_{ij}}{1-q^{-1}z_{ij}}\nn\\
    &=\fft1{N!}\fft{q^N}{(q-1)^N}\int\prod_i\fft{dz_i}{2\pi iz_i}\prod_{i\ne j}\fft{q(1-z_{ij})}{q-z_{ij}}\nn\\
    &=\fft1{N!}\fft{(-1)^Nq^{N^2}}{(1-q)^N}\int\prod_i\fft{dz_i}{2\pi iz_i}\prod_{i\ne j}\fft{1-z_{ij}}{1-qz_{ij}}=(-1)^Nq^{N^2}Z_N(q).
\end{align}
Taking (\ref{eq:ZN(q)}) as the definition of $Z_N(q)$ then gives
\begin{equation}
    Z_m^{(\mathrm{ci})}(q^{-1})=(-1)^m\frac{q^{m^2}}{(q;q)_m}=\frac{q^{m(m-1)/2}}{(q^{-1};q^{-1})_m}\qquad(\mbox{by contour integration}).
\label{eq:CI}
\end{equation}
However, for $m>1$, this is not the same result as analytically continuing (\ref{eq:ZN(q)}), which gives simply
\begin{equation}
    Z_m^{(\mathrm{ac})}(q^{-1})=\fft1{(q^{-1};q^{-1})_m}\qquad(\mbox{by analytic continuation}).
\label{eq:AC}
\end{equation}
Note that both expressions, (\ref{eq:CI}) and (\ref{eq:AC}), are well defined for finite $m$, even with $q^{-1}$ outside the unit circle.

This discrepancy between the contour integral and direct analytic continuation plays an important role in the giant graviton expansion.  In a series expansion of the super-Yang-Mills partition function, the fugacities are expanded around zero, so they lie within the unit circle.  However, when considering the giant graviton contributions, one has to work with wrapped D3-branes with different charge assignments.  Thus the giant graviton index, $\hat Z_m(q)$, is at least formally expressed as an index $Z_m(q^{-1})$ with fugacity outside the unit circle.  It is then an issue how this index is evaluated for fugacities outside the unit circle.  In \cite{Arai:2020qaj,Imamura:2021ytr}, a particular ``pole selection rule'' was proposed so that the contour integral would result in $\hat Z_m(q)=1/(q^{-1};q^{-1})_m$ for the 1/2 BPS index, while in \cite{Gaiotto:2021xce}, Gaiotto and Lee worked directly with the analytic continuation observation, (\ref{eq:AC}), for the giant graviton index.  The issue of discontinuous jumps in the contour integral as poles enter and exit the contour can be thought of as a wall-crossing phenomenon \cite{Lee:2022vig}, and this was investigated in detail in \cite{Beccaria:2023zjw} for the Schur index.

\subsection{Murthy's giant graviton expansion}

As demonstrated in \cite{Eniceicu:2023uvd}, Murthy's giant graviton expansion of the 1/2 BPS index does not directly reproduce the D3-brane expansion in the sense that $G_N^{(m)}(q)\ne q^{mN}\hat Z_m(q)$.  Nevertheless, the two expansions can be related to each other in the sense that $\hat Z_m(q)$ is fully encoded in the first $m$ terms of Murthy's expansion, namely the terms $G_N^{(m')}(q)$ with $1\le m'\le m$.  To explore this connection, we begin with Murthy's giant graviton expression, (\ref{eq:EnGG2}), applied to the 1/2 BPS index.  In this case, the single letter index is $f(q)=q$, so that $\hat f(q)=q/(1-q)=\sum_{k\ge1}q^k$.  Making use of (\ref{eq:PE(x)}) and rearranging some of the $q$-Pochhammer symbols, we find
\begin{equation}
    G_N^{(m)}=\fft1{(m!)^2}(q;q)_\infty^{2m}\int\prod_i\left[\fft{dz_i}{2\pi iz_i}\fft{dw_i}{2\pi iw_i}\left(\fft{w_i}{z_i}\right)^N\right]\fft{\prod_{i\ne j}(\fft{z_i}{z_j};q)_\infty(\fft{w_i}{w_j};q)_\infty}{\prod_{i,j}(\fft{z_i}{w_j};q)_\infty(\fft{w_i}{z_j};q)_\infty}.
\label{eq:Murthy1/2}
\end{equation}
Following \cite{Eniceicu:2023uvd}, we choose the integration contours such that $|q|<|w_i/z_j|<1$ and perform the $w_i$ contour integrals first.

As shown in \cite{Eniceicu:2023uvd}, the poles inside the $w_i$ contours correspond to the zeros of $(z_i/w_j;q)_\infty$ and take the form $w_j=q^kz_i$ for some $i$, $j$ and $k\ge1$.  A complete set of poles are then specified by
\begin{equation}
    w_1=q^{k_1}z_{i_1},\qquad w_2=q^{k_2}z_{i_2},\qquad\ldots,\qquad w_m=q^{k_m}z_{i_m}.
\label{eq:multipoles}
\end{equation}
To proceed, note that, if $i_1=i_2$, then $w_1/w_2=q^{k_1-k_2}$.  Since this is an integer power of $q$, either $(w_1/w_2;q)_\infty$ or $(w_2/w_1;q)_\infty$ will vanish.  Since this applies to any pair of $w_i$'s, we see that whenever any pair of $z_i$'s are equal in (\ref{eq:multipoles}), the residue will vanish.  As a result, we can restrict to the poles
\begin{equation}
    w_i=q^{k_i}z_i,\qquad k_i\ge1,
\label{eq:1/2BPSwpoles}
\end{equation}
up to an overall factor of $m!$ accounting for the permutations of the $z_i$'s.

Summing over the resides of the poles given by (\ref{eq:1/2BPSwpoles}), and using the $q$-Pochhammer identities
\begin{align}
    \fft{(a;q)_\infty(a^{-1}q;q)_\infty}{(aq^k;q)_\infty(a^{-1}q^{1-k};q)_\infty}&=(-1)^ka^kq^{k(k-1)/2}\qquad(a\ne1),\nn\\
    \fft{(q;q)_\infty^2}{(q^k;q)_\infty(q^{1-k};q)'_\infty}&=(-1)^{k-1}q^{k(k-1)/2},
\label{eq:pochids}
\end{align}
Murthy's expression becomes
\begin{align}
    G_N^{(m)}(q)&=\sum_{K\ge m}q^{KN}\fft{(-1)^{K-m}}{m!}q^{K(K-m)/2}\kern-.5em\sum_{\genfrac{}{}{0pt}{1}{k_i\ge1}{\sum_ik_i=K}}\Biggl[\prod_i\fft1{1-q^{-k_i}}\nn\\
    &\kern8em\times\int\prod_i\fft{dz_i}{2\pi iz_i}\prod_{i\ne j}\fft{1-z_{ij}}{1-z_{ij}q^{-k_j}}\prod_{i<j}\fft{q^{(k_i-k_j)/2}z_i-q^{(k_j-k_i)/2}z_j}{z_i-z_j}\Biggr].
\label{eq:1/2BPSzint}
\end{align}
Note that the only $N$-dependent term is the $q^{KN}$ pre-factor.  This demonstrates that Murthy's $G_N^{(m)}(q)$ contributes not just to the D3-brane giant graviton index, $\hat Z_m(q)$, but to all such indices with $m$ or more giant gravitons.  Equivalently, the D3-brane index $\hat Z_K(q)$ will receive contributions for $G_N^{(m)}(q)$ with $m=1,2,\ldots,K$ \cite{Eniceicu:2023uvd}.

The simplest case corresponds to a single giant graviton, where $\hat Z_1(q)$ will only receive a contribution from the $K=1$ term in $G_N^{(1)}(q)$.  We can see this explicitly by setting $m=1$ and $K=k_1=1$ in (\ref{eq:1/2BPSzint}).  In this case, we can easily verify that
\begin{equation}
    \left.G_N^{(1)}(q)\right|_{K=1}=\frac{q^N}{1-q^{-1}}\int\frac{dz_1}{2\pi iz_1}=\frac{q^N}{1-q^{-1}}=\frac{q^N}{(q^{-1}; q^{-1})_1}=q^N\hat Z_{(1)}(q),
\end{equation}
where $\hat Z_m(q)$ is given in (\ref{eq:hatZ_m(q)}).  Of course, $G_N^{(m)}(q)$ receives contributions from $K>1$ as well.  The expansion is straightforward
\begin{align}
    G_N^{(1)}(q)&=\sum_{K\ge1}q^{KN}(-1)^{K-1}\frac{q^{K(K-1)/2}}{1-q^{-K}}\nn\\
    &=q^N\frac1{1-q^{-1}}-q^{2N}\frac{q}{1-q^{-2}}+q^{3N}\frac{q^3}{1-q^{-3}}-\cdots.
\label{eq:GN1expand}
\end{align}
Note that the $q^{2N}$ term does not match the two giant graviton expression, $\hat Z_2(q)$, since $\hat Z_2(q)$ will also pick up a term from $G_N^{(2)}(q)$.  We can see this explicitly by evaluating (\ref{eq:1/2BPSzint}) with $m=2$.  However, before doing so, it is interesting to examine the structure of the $K=m$ term in $G_N^{(m)}(q)$.

Since the integral form of Murthy's giant graviton expression, (\ref{eq:1/2BPSzint}), is expressed as a sum first over $K$ and then over partitions of $K$ into positive integers, we introduce the notation
\begin{equation}
    G_N^{(m)}(q)=\sum_{K\ge m}q^{NK}\tilde G_N^{(m;K)}(q),
\end{equation}
where
\begin{align}
    \tilde G_N^{(m;K)}(q)&=\fft{(-1)^{K-m}}{m!}q^{K(K-m)/2}\kern-.5em\sum_{\genfrac{}{}{0pt}{1}{k_i\ge1}{\sum_ik_i=K}}\Biggl[\prod_i\fft1{1-q^{-k_i}}\nn\\
    &\kern6em\times\int\prod_i\fft{dz_i}{2\pi iz_i}\prod_{i\ne j}\fft{1-z_{ij}}{1-z_{ij}q^{-k_j}}\prod_{i<j}\fft{q^{(k_i-k_j)/2}z_i-q^{(k_j-k_i)/2}z_j}{z_i-z_j}\Biggr].
\label{eq:tildeGNmK}
\end{align}
Then the connection between the D3-brane expansion and Murthy's expansion is
\begin{equation}
    \hat Z_K(q)=\sum_{m=1}^K\tilde G_N^{(m;K)}(q).
\label{eq:GintoZhat}
\end{equation}
Note that the top of this sum, with $m=K$ is particularly simple since this requires $k_i=1$, resulting in the expression
\begin{equation}
    \tilde G_N^{(K;K)}(q)=\fft1{m!}\fft1{(1-q^{-1})^K}\int\prod_i\fft{dz_i}{2\pi iz_i}\prod_{i\ne j}\fft{1-z_{ij}}{1-z_{ij}q^{-1}}.
\label{eq:ieq-1}
\end{equation}
Comparing with the original integral expression for the 1/2 BPS index, (\ref{eq:PE1/2}), we see that
\begin{equation}
     \tilde G_N^{(K;K)}(q)=Z_K^{(\mathrm{ci})}(q^{-1})\qquad(\mbox{by contour integration}),
\end{equation}
which is suggestive of Gaiotto and Lee's observation, (\ref{eq:GLGGE}).  However, as discussed above, the contour integral expression for $Z_K(q^{-1})$ with $|q|<1$ does not match the analytic continuation result, (\ref{eq:AC}), used in defining the D3-brane giant graviton index, $\hat Z_K(q)$.

Splitting off the $m=K$ term in (\ref{eq:GintoZhat}), and making use of the contour integration result, (\ref{eq:CI}), we can write
\begin{equation}
    \hat Z_K(q)=\fft{q^{K(K-1)/2}}{(q^{-1};q^{-1})_K}+\sum_{m=1}^{K-1}\tilde G_N^{(m;K)}(q).
\label{eq:ZKfromGtilde}
\end{equation}
For $K=1$, this is trivial.  But for $K=2$, we have
\begin{equation}
    \hat Z_2(q)=\fft{q}{(q^{-1};q^{-1})_2}+\tilde G_N^{(1;2)}(q).
\end{equation}
From (\ref{eq:GN1expand}), we find
\begin{equation}
    \tilde G_N^{(1;2)}(q)=-\fft{q}{1-q^{-2}}=\fft{1-q}{(q^{-1};q^{-1})_2}.
\end{equation}
As a result, we have
\begin{equation}
    \hat Z_2(q)=\fft{q}{(q^{-1};q^{-1})_2}+\fft{1-q}{(q^{-1};q^{-1})_2}=\fft1{(q^{-1};q^{-1})_2},
\end{equation}
which agrees with the D3-brane giant graviton index, (\ref{eq:hatZ_m(q)}).

This pattern continues for an arbitrary number of giant gravitons.  In particular, in \cite{Eniceicu:2023uvd}, Eniceicu derived a general expression for $\tilde G_N^{(m;K)}(q)$ in terms of a recurrence relation by explicitly performing the $z_i$ integrals in (\ref{eq:tildeGNmK}) with $q$ inside the unit circle.  He was then able to sum up Murthy's giants and explicitly verify the relation, (\ref{eq:ZKfromGtilde}), between the two giant graviton expansions.

What we have shown is that, after evaluation of the $w_i$ contour integrals in Murthy's expression, (\ref{eq:Murthy1/2}), we arrive at the expansion, (\ref{eq:ZKfromGtilde}), which is of the form
\begin{equation}
    \hat Z_k(q)=\sum_{m=1}^k\tilde G_N^{(m;k)}(q)=Z_k^{(\mathrm{ci})}(q^{-1})+\cdots,
\end{equation}
where the (ci) superscript is a reminder that evaluation of the 1/2 BPS partition function, (\ref{eq:PE1/2}), is to be done by contour integration.  While this is superficially similar to the D3-brane result, (\ref{eq:GLGGE}), the results actually differ because of pole selection issues as discussed above.  As we will see below, these similarities and differences persist when considering reduced supersymmetry indices.

\section{The Schur index}
\label{sec:Schur}

Having clarified the connection between the D3-brane giants and Murthy's giants for the 1/2 BPS index, we now turn to the Schur index.  In the symmetric formulation, the 1/8 BPS index is a function of two fugacities, which we denote $x$ and $y$.  The single letter index is
\begin{equation}
    f(x,y)=1-\fft{(1-x)(1-y)}{1-xy},
\label{eq:f(x,y)}
\end{equation}
which leads to the integral expression for the full index
\begin{align}
    Z_N(x,y)=\fft1{N!}\fft{(xy;xy)_\infty^{2N}}{(x;xy)_\infty^N(y;xy)_\infty^N}\int\prod_i\fft{dz_i}{2\pi iz_i}\prod_{i\ne j}\fft{(z_{ij};xy)_\infty(xyz_{ij};xy)_\infty}{(xz_{ij};xy)_\infty(yz_{ij};xy)_\infty}.
\label{eq:N=4Schur}
\end{align}
With two equal fugacities, $q=x=y$, the exact unflavored Schur index, $Z_N(q)$, was obtained in \cite{Bourdier:2015wda} as an expansion
\begin{equation}
    Z_N(q)=Z_\infty(q)\sum_{k\ge0}(-1)^k\left[\binom{N+k}N+\binom{N+k-1}N\right]q^{kN+k^2},
\label{eq:BDFgg}
\end{equation}
where the infinite-$N$ index is given by
\begin{equation}
    Z_\infty(q)=\fft1{(q;q)_\infty(q;q^2)_\infty}.
\label{eq:Schurqinf}
\end{equation}
In fact, it was noted in \cite{Bourdier:2015wda} that (\ref{eq:BDFgg}) directly yields a giant graviton expansion with
\begin{equation}
    \hat Z_k(q)=(-1)^k\left[\binom{N+k}{N}+\binom{N+k-1}{N}\right]q^{k^2}=(-1)^k\fft{N+2k}k\binom{N+k-1}Nq^{k^2}.
\label{eq:ZhatBDF}
\end{equation}
However, the structure of the flavored Schur index with independent fugacities $x$ and $y$ is more intricate.

\subsection{The D3-brane giant graviton expansion}

The giant graviton expansion of the flavored Schur index was considered in \cite{Arai:2020qaj}, where it was conjectured to take the form
\begin{equation}
    \fft{Z_N(x,y)}{Z_\infty(x,y)}=\sum_{(m_1,m_2)\ge0} x^{m_1N}y^{m_2N}\hat Z_{(m_1,m_2)}(x,y),
\label{eq:D3Schur}
\end{equation}
with
\begin{equation}
    Z_\infty(x,y)=\fft{(xy;xy)_\infty}{(x;x)_\infty(y;y)_\infty},
\end{equation}
generalizing (\ref{eq:Schurqinf}).  Here $\hat Z_{(m_1,m_2)}$ is the index for $m_1$ D3-branes wrapped on $X=0$ and $m_2$ D3-branes wrapped on $Y=0$.  The $(m_1,m_2)$ D3-brane system can be described by a two-node quiver, and $\hat Z_{(m_1,m_2)}$ takes the form
\begin{equation}
    \hat Z_{(m_1,m_2)}(x,y)=Z_{m_1}(x^{-1},xy)(xy)^{m_1m_2}Z_{m_2}(xy,y^{-1}),
\label{eq:afim}
\end{equation}  
where $Z_m(x,y)$ is the $\mathcal N=4$ SYM index, (\ref{eq:N=4Schur}), on a stack of $m$ D3-branes.  The factor $(xy)^{m_1m_2}$ is the contribution to the index from the bifundamental hypermultiplet corresponding to strings stretched between the two stacks of D3-branes.

The evaluation of (\ref{eq:afim}) requires care because of potential analytic continuation issues related to the inverse fugacities, $x^{-1}$ and $y^{-1}$.  The expansion of \cite{Arai:2020qaj} followed the pole selection rule subsequently discussed in \cite{Imamura:2021ytr} and given a more precise treatment and brane interpretation in \cite{Lee:2022vig}.  The symmetric $Z_N(x,y)$ form of the Schur index was extensively investigated in \cite{Beccaria:2023zjw}, where it was also connected to the asymmetric $Z_N(x,q/x)$ Gaiotto-Lee expansion \cite{Gaiotto:2021xce}.  Exact results for the brane indices have been obtained in \cite{Beccaria:2024szi}.

\subsection{Murthy's giant graviton expansion}
\label{sec:murthyschur}

As in the 1/2 BPS case, we wish to recover the D3-brane giant graviton expansion, (\ref{eq:afim}), from Murthy's expansion of, (\ref{eq:EnGG2}).  Starting with the single letter Schur index, (\ref{eq:f(x,y)}), we find
\begin{equation}
    \hat f(x,y)\equiv\fft{f(x,y)}{1-f(x,y)}=\fft{x}{1-x}+\fft{y}{1-y}=\sum_{k\ge1}x^k+y^k.
\end{equation}
As a result, the expression for $G_N^{(m)}$ can be written as
\begin{align}
    G_N^{(m)}(x,y)&=\fft{(-1)^m}{(m!)^2}(x;x)_\infty^{2m}(y;y)_\infty^{2m}\int\prod_i\left[\fft{dz_i}{2\pi iz_i}\fft{dw_i}{2\pi iw_i}\left(\fft{w_i}{z_i}\right)^{N+m}\right]\nn\\
    &\kern10em\times\fft{\prod_{i\ne j}(z_{ij};x)_\infty(w_{ij};x)_\infty(z_{ij}y;y)_\infty(w_{ij}y;y)_\infty}{\prod_{i,j}(\fft{z_i}{w_j}x;x)_\infty(\fft{w_i}{z_j};x)_\infty(\fft{z_i}{w_j}y;y)_\infty(\fft{w_i}{z_j};y)_\infty}.
\label{eq:SchurGNm}
\end{align}
If we reduce to the unflavored index by setting $q=x=y$, we will end up with
\begin{align}
    G_N^{(m)}(q)&=\fft{(-1)^m}{(m!)^2}(q;q)_\infty^{4m}\int\prod_i\left[\fft{dz_i}{2\pi iz_i}\fft{dw_i}{2\pi iw_i}\left(\fft{w_i}{z_i}\right)^{N+m}\right]\nn\\
    &\kern8em\times\fft{\prod_{i\ne j}(z_{ij};q)_\infty(z_{ij}q;q)_\infty(w_{ij};q)_\infty(w_{ij}q;q)_\infty}{\prod_{i,j}(\fft{z_i}{w_j}q;q)_\infty^2(\fft{w_i}{z_j};q)_\infty^2}.
\end{align}
As noted in \cite{Eniceicu:2023uvd}, this expression has double poles in the integrand.  Obtaining the residue at a double pole will involve taking the first derivative of the remaining integrand, and hence will potentially contribute a factor of $N$ from the $(w_i/z_i)^{N+m}$ factor.  Assuming we perform the $w_i$ integrals first, since there are as many double pole residues as the number of $w_i$ integrals, one can potentially end up with an order-$m$ polynomial of $N$ in the expansion of $G_N^{(m)}(q)$, which is consistent with the D3-brane index $\hat Z_k(q)$ in (\ref{eq:ZhatBDF}) being a polynomial of degree $k$ in $N$.

In the general flavored case, (\ref{eq:SchurGNm}) only has simple poles, so the procedure of \cite{Eniceicu:2023uvd} to first integrate over $w_i$ and then over $z_i$ can be carried out without the additional difficulty of higher order poles.  In this case, we do not expect any additional $N$ dependence in $G_N^{(m)}$ beyond the prefactors of the form $x^N$ or $y^N$ raised to appropriate powers.

\subsubsection{Performing the $w_i$ integrals}

As in the 1/2 BPS case, we proceed by performing the $w_i$ integrals, followed by the $z_i$ integrals.  Examining Murthy's expression for the Schur index, (\ref{eq:SchurGNm}), we see that the $w_i$ poles occur at either $w_i=z_jx^k$ or $w_i=z_jy^k$.  We thus have
\begin{equation}
    w_1=a_1^{k_1}z_{i_1},\qquad w_2=a_2^{k_2}z_{i_2},\qquad\cdots,\qquad w_m=a_m^{k_m}z_{i_m},
\label{eq:multiwres}
\end{equation}
with $k_i\ge1$ and where each $a_i$ can be either $x$ or $y$.  In the 1/2 BPS case, all the $z_{i_n}$'s have to be unique, otherwise the numerator factor and hence the residue vanishes.  Here, however, consider, \textit{e.g.}, $w_1/w_2=a_1^{k_1}a_2^{-k_2}z_{i_1}/z_{i_2}$.  If $i_1=i_2$, this reduces to  $w_1/w_2=a_1^{k_1}a_2^{-k_2}$.  If $a_1=a_2=x$ or $a_2=a_2=y$ then once again one of the $q$-Pochhammers in the numerator will vanish.  However, if $a_1$ and $a_2$ are distinct (with one being $x$ and the other $y$), then there is no longer any reason for the residue to vanish.  Thus, for the Schur index, the various $z_{i_n}$ factors in (\ref{eq:multiwres} do not have to be unique.  However, each individual $z_{i_n}$ can appear at most twice (once with a $x^{k_i}$ factor and once with a $y^{k_j}$ factor).

Changing the notation slightly, we assume there are $p$ pairs of $w_i$'s that involve the same $z_{i_n}$.  The remaining $w_i$'s are then unpaired, with $n_1$ of them corresponding to powers of $x$ and $n_2$ of them corresponding to powers of $y$.  The number of Murthy's giants is then $m=2p+n_1+n_2$.  Up to permutations, the poles are then given by
\begin{align}
    &\begin{aligned}w_{2a-1}&=x^{\ell_a}z_{2a-1}\\ w_{2a}&=y^{\tilde\ell_a}z_{2a-1}\end{aligned}&& (a=1,\ldots,p)\nn\\
    &\kern.1em w_{2p+i}=x^{k_i}z_{2p+i}&&(i=1,\ldots,n_1)\nn\\
    &\kern-1.3em w_{2p+n_1+\tilde i}=y^{\tilde k_{\tilde i}}z_{2p+n_1+\tilde i}&& (\tilde i=1\ldots,n_2),
\label{eq:Schurpoles}
\end{align}
and are specified by a set of positive integers $\{\ell_a,\tilde\ell_a,k_i,\tilde k_{\tilde i}\}$.  The degeneracy of such a given combination involves putting the $w_i$'s and $z_i$'s into the four categories above along with permutations within each category
\begin{equation}
    \mathrm{degeneracy}=\left(\fft{m!}{p!^2n_1!n_2!}\right)^2p!^2n_1!n_2!=\frac{m!^2}{p!^2n_1!n_2!}.
\end{equation}

After performing the $w_i$ integrals, we end up with an expression for $G_N^{(m)}$ which is a sum over the set of poles (\ref{eq:Schurpoles}) of the form
\begin{equation}
    G_N^{(m)}(x,y)=\sum_{\{\ell_a,\tilde \ell_a,k_i,\tilde k_{\tilde i}\}>0}G_N^{(\ell_a,\tilde \ell_a,k_i,\tilde k_{\tilde i})}(x,y),
\label{eq:GMmsum}
\end{equation}
where
\begin{align}
    G_N^{(\ell_a,\tilde \ell_a,k_i,\tilde k_{\tilde i})}(x,y)&=\fft{(-1)^{\ell+\tilde\ell+k+\tilde k+(p+1)(m+1)+1}}{p!^2n_1!n_2!}(x^{k+\ell}y^{\tilde k+\tilde\ell})^N(xy)^{(k+\ell)(\tilde k+\tilde\ell)}\nn\\
    &\quad\int\prod_a\fft{dz_{2a-1}}{2\pi iz_{2a-1}}\fft{dz_{2a}}{2\pi iz_{2a}}\prod_i\fft{dz_i}{2\pi iz_i}\prod_{\tilde i}\fft{dz_{\tilde i}}{2\pi iz_{\tilde i}}\nn\\
    &\quad\times\prod_a\left[\left(\fft{z_{2a-1}}{z_{2a}}\right)^{N+\ell+\tilde\ell+k+\tilde k-p}\prod_i\fft{z_i}{z_{2a-1}}\prod_{\tilde i}\fft{z_{\tilde i}}{z_{2a-1}}\right]I_{\mathrm{pairs}}I_{\mathrm{mixed}}I_{\mathrm{singles}}.
\label{eq:GMStot}
\end{align}
We have broken up the integrand into three components
\begin{align}
    I_{\mathrm{pairs}}&=\prod_{a\ne b}(1-\frac{z_{2a}}{z_{2b}})\prod_{a<b}x^{(\ell_b-\ell_a)/2}y^{(\tilde\ell_a-\tilde\ell_b)/2}(1-\frac{z_{2a-1}}{z_{2b-1}}x^{\ell_a-\ell_b})(1-\frac{z_{2b-1}}{z_{2a-1}}y^{\tilde\ell_b-\tilde\ell_a})\nn\\
    &\quad\times x^{\ell(\ell-p)/2}\prod_{a,b}\frac{(\frac{z_{2a}}{z_{2b}}y;y)_\infty(\frac{z_{2a-1}}{z_{2b-1}}y^{\ell_a-\ell_b}y;y)_\infty}{(\frac{z_{2a-1}}{z_{2b}}x^{\ell_a}y;y)_\infty(\frac{z_{2a}}{z_{2b-1}}x^{-\ell_b};y)_\infty}\nn\\
    &\quad\times y^{\tilde\ell(\tilde\ell-p)/2}\prod_{a,b}\frac{(\frac{z_{2a}}{z_{2b}}x;x)_\infty(\frac{z_{2a-1}}{z_{2b-1}}y^{\tilde\ell_a-\tilde\ell_b}x;x)_\infty}{(\frac{z_{2a-1}}{z_{2b}}y^{\tilde\ell_a}x;x)_\infty(\frac{z_{2a}}{z_{2b-1}}y^{-\tilde\ell_b};x)_\infty},
\label{eq:pairs}
\end{align}
\begin{align}
    I_{\mathrm{mixed}}&=x^{(k-n_1)\ell}\prod_{a,i}\frac{(\fft{z_{2a-1}}{z_i}x^{\ell_a-k_i};y)_\infty(\fft{z_i}{z_{2a-1}}x^{-\ell_a+k_i}y;y)_\infty(\fft{z_{2a}}{z_i};y)_\infty(\fft{z_i}{z_{2a}}y;y)_\infty}{(\fft{z_i}{z_{2a-1}}x^{-\ell_a};y)_\infty(\fft{z_{2a-1}}{z_i}x^{\ell_a}y;y)_\infty(\fft{z_i}{z_{2a}}x^{k_i}y;y)_\infty(\fft{z_{2a}}{z_i}x^{-k_i};y)_\infty}\nn\\
    &\quad\times y^{(\tilde k-n_2)\tilde\ell}\prod_{a,\tilde i}\frac{(\fft{z_{2a-1}}{z_{\tilde i}}y^{\tilde\ell_a-\tilde k_{\tilde i}};x)_\infty(\fft{z_{\tilde i}}{z_{2a-1}}y^{-\tilde\ell_a+\tilde k_{\tilde i}}x;x)_\infty(\fft{z_{2a}}{z_{\tilde i}};x)_\infty(\fft{z_{\tilde i}}{z_{2a}}x;x)_\infty}{(\fft{z_{\tilde i}}{z_{2a-1}}y^{-\tilde\ell_a};x)_\infty(\fft{z_{2a-1}}{z_{\tilde i}}y^{\tilde\ell_a}x;x)_\infty(\fft{z_{\tilde i}}{z_{2a}}y^{\tilde k_{\tilde i}}x;x)_\infty(\fft{z_{2a}}{z_{\tilde i}}y^{-\tilde k_{\tilde i}};x)_\infty},
\end{align}
and
\begin{align}
    I_{\mathrm{singles}}&=x^{k(k-1)/2}\prod_{i<j}x^{-k_i}(1-z_{ij}x^{k_i-k_j})(1-z_{ji})\prod_{i,j}\fft{(z_{ij}x^{k_i-k_j}y;y)_\infty(z_{ij}y;y)_\infty}{(z_{ij}x^{k_i}y;y)_\infty(z_{ij}x^{-k_j};y)_\infty}\nn\\
    &\quad\times y^{\tilde k(\tilde k-1)/2}\prod_{\tilde i<\tilde j}y^{-\tilde k_{\tilde i}}(1-z_{\tilde i\tilde j}y^{\tilde k_{\tilde i}-\tilde k_{\tilde j}})(1-z_{\tilde j\tilde i})\prod_{\tilde i,\tilde j}\fft{(z_{\tilde i\tilde j}y^{\tilde k_{\tilde i}-\tilde k_{\tilde j}}x;x)_\infty(z_{\tilde i\tilde j}x;x)_\infty}{(z_{\tilde i\tilde j}y^{\tilde k_{\tilde i}}x;x)_\infty(z_{\tilde i\tilde j}y^{-\tilde k_{\tilde j}};x)_\infty}.
\end{align}
Note that we have used the shorthand notation $k=\sum_ik_i$, etc., along with $z_{2p+i}\to z_i$ and $z_{2p+n_1+\tilde i}\to z_{\tilde i}$.  These expressions were obtained by using the $q$-Pochhammer identities, (\ref{eq:pochids}), on the residues.

In order to break this down into wrapped D3-brane contributions, we note that the rank dependence of Murthy's expansion arises from the factors in (\ref{eq:GMStot}) that depend on $N$
\begin{equation}
    (x^{k+\ell}y^{\tilde k+\tilde\ell})^N\prod_a\left(\fft{z_{2a-1}}{z_{2a}}\right)^N.
\end{equation}
While the prefactor is easy to understand, the $z_{2a-1}/z_{2a}$ dependent terms do not have a definite giant graviton contribution, and can only be fully determined after performing the $z_{2a-1}$ and $z_{2a}$ integrals.  However, since the poles will involve powers of $x$ and $y$, we see that the $N$-dependence will only arise as powers of $x$ and $y$.  While it is not manifestly obvious that only non-negative powers will arise, consistency with the D3-brane expression, (\ref{eq:D3Schur}), will demand this to be the case.

\subsubsection{The zero-pair contributions}

In principle, the next step is to perform the $z_i$ integrals.  However, the contribution of poles and residues is quite involved.  Furthermore, when performing the contour integrals, one will also have to consider which poles lie within the contours when evaluating the multivariate residues.  This will depend on the magnitudes of $x$ and $y$ and ratios of their powers.  We will not be complete here but instead will focus on some general observations.

For the Schur index, the set $w_i$ poles that come in pairs do not have an obvious 1/2 BPS counterpart.  However, for $p=0$, the integral expression for $G_N^{(m)}(x,y)$ simplifies considerably.  We thus write
\begin{equation}
    G_N^{(m)}(x,y)=\sum_{\{k_i,\tilde k_{\tilde i}\}>0}G_N^{(p=0;k_i,\tilde k_{\tilde i})}(x,y)+\sum_{\genfrac{}{}{0pt}{1}{p>0}{\{\ell_a,\tilde \ell_a,k_i,\tilde k_{\tilde i}\}>0}}G_N^{(\ell_a,\tilde\ell_a,k_i,\tilde k_{\tilde i})}(x,y),
\end{equation}
where the $p=0$ contribution takes the form
\begin{align}
    G_N^{(p=0;k_i,\tilde k_{\tilde i})}(x,y)&=\fft{(-1)^{(k-n_1)+(\tilde k-n_2)}}{n_1!n_2!}(x^ky^{\tilde k})^N(xy)^{k\tilde k}\int\prod_i\fft{dz_i}{2\pi iz_i}\prod_{\tilde i}\fft{dz_{\tilde i}}{2\pi iz_{\tilde i}}\nn\\
    &\quad\times x^{k(k-1)/2}\prod_{i<j}x^{-k_i}(1-z_{ij}x^{k_i-k_j})(1-z_{ji})\prod_{i,j}\fft{(z_{ij}x^{k_i-k_j}y;y)_\infty(z_{ij}y;y)_\infty}{(z_{ij}x^{k_i}y;y)_\infty(z_{ij}x^{-k_j};y)_\infty}\nn\\
    &\quad\times y^{\tilde k(\tilde k-1)/2}\prod_{\tilde i<\tilde j}y^{-\tilde k_{\tilde i}}(1-z_{\tilde i\tilde j}y^{\tilde k_{\tilde i}-\tilde k_{\tilde j}})(1-z_{\tilde j\tilde i})\prod_{\tilde i,\tilde j}\fft{(z_{\tilde i\tilde j}y^{\tilde k_{\tilde i}-\tilde k_{\tilde j}}x;x)_\infty(z_{\tilde i\tilde j}x;x)_\infty}{(z_{\tilde i\tilde j}y^{\tilde k_{\tilde i}}x;x)_\infty(z_{\tilde i\tilde j}y^{-\tilde k_{\tilde j}};x)_\infty}.
\end{align}
Note that this has the factorized form
\begin{equation}
     G_N^{(p=0;k_i,\tilde k_{\tilde i})}(x,y)=(x^ky^{\tilde k})^N F_{\{k_i\}}(x,y)(xy)^{k\tilde k}F_{\{\tilde k_{\tilde i}\}}(y,x),
\label{eq:GNmsimpGGE}
\end{equation}
where
\begin{align}
    F_{\{k_i\}}(x,y)&=\fft{(-1)^{k-n}}{n!}x^{k(k-1)/2}\int\prod_i\fft{dz_i}{2\pi iz_i}\nn\\
    &\qquad\times\prod_{i<j}x^{-k_i}\left(1-z_{ij}x^{k_i-k_j}\right)\left(1-z_{ji}\right)\prod_{i,j}\fft{(z_{ij}x^{k_i-k_j}y;y)_\infty(z_{ij}y;y)_\infty}{(z_{ij}x^{k_i}y;y)_\infty(z_{ij}x^{-k_j};y)_\infty}.
\label{eq:FxyGGE}
\end{align}
The full contribution to the index involves a sum over all poles.  For the $p=0$ non-paired pole contributions, we still have to sum over all sets of positive integers $\{k_i,\tilde k_{\tilde i}\}$.  However, the lowest order contribution, with $k_i=1$ and $\tilde k_{\tilde i}=1$ is particularly interesting, as it reduces to the D3-brane giant graviton expansion of \cite{Arai:2020qaj}
\begin{equation}
    G_N^{(p=0;\mathbf 1,\mathbf 1)}(x,y)=(x^{n_1}y^{n_2})^NZ_{n_1}(x^{-1},xy)(xy)^{n_1n_2}Z_{n_2}(xy,y^{-1}),
\label{eq:murthytoafim}
\end{equation}
where $Z_n(x,y)$ is the integral expression for the $U(n)$ Schur index given in (\ref{eq:N=4Schur}).

It is important to realize, however, that for Murthy's giant graviton expansion, the contour integrals in the integral expression, (\ref{eq:N=4Schur}), are performed with the $z_i$'s on the unit circle and with a conventional treatment of the poles.  This in particular differs from the analytic continuation (\textit{i.e.}\ pole selection rule) involved in defining the D3-brane index, $\hat Z_{(n_1,n_2)}$ defined in (\ref{eq:afim}).  In the 1/2-BPS case, we were able to evaluate the analogous $Z_m(q^{-1})$ explicitly by both analytic continuation and direct contour integration
\begin{equation}
    Z_m(q)=\fft1{(q;q)_m}\qquad\Rightarrow\qquad Z_m(q^{-1})=\begin{cases}\fft1{(q^{-1};q^{-1})_m},&\mbox{analytic continuation};\\[2mm]\fft{q^{m(m-1)/2}}{(q^{-1};q^{-1})_m},&\mbox{contour integration}.\end{cases}
\end{equation}
Unfortunately, we are not aware of a similar compact expression for the Schur index.  Nevertheless, while schematically (\ref{eq:murthytoafim}) appears to derive the D3-brane giant graviton index from Murthy's expansion, just as in the 1/2-BPS case the reality is more complicated.  In particular, the contour integration result for $G_N^{(p=0;\mathbf1,\mathbf1)}$ in itself does not reproduce the giant graviton index $\hat Z_{(n_1,n_2)}$.  The missing terms come from poles with $k_i>1$ and/or $\tilde k_{\tilde i}>1$ as well as from contributions with $z_i$ pairs (\textit{i.e.}\ with $p>0$).  While we have been unable to derive a general relation between Murthy's $G_N^{(m)}$ and the D3-brane index $\hat Z_{(n_1,n_2)}$, it is nevertheless instructive to see how this works for the case of one and two giant gravitons.

\subsection{One giant graviton}

For a single giant graviton, one cannot have any $w_i$ pairs.  Thus the giant graviton contribution can be obtained directly from the $p=0$ expression (\ref{eq:GNmsimpGGE}) with $(n_1,n_2)=(1,0)$ and $(0,1)$
\begin{equation}
    G_N^{(1)}(x,y)=\sum_{k>0}x^{kN}F_k(x,y)+y^{kN}F_k(y,x),
\end{equation}
where
\begin{equation}
    F_k(x,y)=(-1)^{k-1}x^{k(k-1)/2}\fft{(y;y)_\infty^2}{(x^{-k};y)_\infty(x^ky;y)_\infty}.
\end{equation}
Note that the integral over $dz_1$ is trivial in this case.  As is by now obvious, Murthy's $G_N^{(1)}$ contributes not only to the single D3-brane index, but to all indices of the form $\hat Z_{(k,0)}$ and $\hat Z_{(0,k)}$.

Splitting off the $k=1$ term, we can write
\begin{align}
    G_N^{(1)}(x,y)&=x^NZ_1(x^{-1},xy)+y^NZ_1(xy,y^{-1})\nn\\
    &\quad+\sum_{k>1}(-1)^{k-1}\left(x^{kN}x^{k(k-1)/2}Z_1(x^{-k},x^ky)+y^{kN}y^{k(k-1)/2}Z_1(y^kx,y^{-k})\right)\nn\\
    &=x^N\hat Z_{(1,0)}(x,y)+y^N\hat Z_{(0,1)}(y,x)+\cdots.
\label{eq:Murthyone}
\end{align}
For this case, since no contour integration is needed, there is no ambiguity in the identification $\hat Z_{(1,0)}(x,y)=Z_1(x^{-1},xy)$ for one giant graviton \cite{Arai:2020qaj,Beccaria:2023zjw} where
\begin{equation}
    Z_1(x,y)=\fft{(xy;xy)_\infty^2}{(x;xy)_\infty(y;xy)_\infty}.
\label{eq:zeeone}
\end{equation}
We thus immediately observe that the single D3-brane index in the expansion (\ref{eq:D3Schur}) is completely contained in Murthy's $G_N^{(1)}$.  From (\ref{eq:Murthyone}), we can also identify the part that contributes to two giant gravitons
\begin{equation}
    G_N^{(1)}=\cdots-x^{2N+1}Z_1(x^{-2},x^2y)-y^{2N+1}Z_1(y^2x,y^{-2})+\cdots.
\label{eq:GN1to2gg}
\end{equation}
The remaining contributions to two giant gravitons originate from $G_N^{(2)}$, which we turn to next.

\subsection{Two giant gravitons}
\label{sec:2gg}

Before looking at $G_N^{(2)}$, we recall from (\ref{eq:D3Schur}) and (\ref{eq:afim}) that there are three individual contributions at the two-giant graviton level
\begin{equation}
    \fft{Z_N(x,y)}{Z_\infty(x,y)}=\cdots+x^{2N}\hat Z_{(2,0)}(x,y)+x^Ny^N\hat Z_{(1,1)}(x,y)+y^{2N}\hat Z_{(0,2)}(x,y)+\cdots,
\end{equation}
where
\begin{align}
    \hat Z_{(2,0)}(x,y)&=Z_2(x^{-1},xy),\qquad Z_{(0,2)}(x,y)=Z_2(xy,y^{-1}),\nn\\
    \hat Z_{(1,1)}(x,y)&=xyZ_1(x^{-1},xy)Z_1(xy,y^{-1}).
\label{eq:hatZ2s}   
\end{align}
Here the Schur index, $Z_N(\cdot ,\cdot )$, is computed from (\ref{eq:N=4Schur}) by analytic continuation.  We would like to see how these contributions arise from a combination of Murthy's $G_N^{(1)}$ and $G_N^{(2)}$.

The contribution of $G_N^{(1)}$ to the two giant graviton index is given in (\ref{eq:GN1to2gg}).  Note that this only contributes to $\hat Z_{(2,0)}(xy)$ and $\hat Z_{(0,2)}(x,y)$.  Moreover, there is no ambiguity in $Z_1(\cdot,\cdot)$, so the $G_N^{(1)}$ contribution is well defined as given.

For $G_N^{(2)}$, there are two possibilities to consider corresponding to zero pairs ($p=0$) and one pair ($p=1$) in (\ref{eq:GMStot})
\begin{equation}
    G_N^{(2)}(x,y)=\sum\left( G_N^{(p=0;k_1,k_2)}(x,y)+G_N^{(p=0;\tilde k_1,\tilde k_2)}+G_N^{(p=0;k,\tilde k)}(x,y)+G_N^{(\ell,\tilde\ell)}(x,y)\right).
\end{equation}
Starting with zero pairs, we have from (\ref{eq:GNmsimpGGE})
\begin{align}
    G_N^{(p=0;k_1,k_2)}(x,y)&=x^{(k_1+k_2)N}F_{(k_1,k_2)}(x,y),\nn\\
    G_N^{(p=0;\tilde k_1,\tilde k_2)}(x,y)&=y^{(\tilde k_1+\tilde k_2)N}F_{(\tilde k_1,\tilde k_2)}(y,x),\nn\\
    G_N^{(p=0;k,\tilde k)}(x,y)&=(x^ky^{\tilde k})^{N+1}F_k(x,y)F_{\tilde k}(y,x),
\end{align}
where $F_{\{k_i\}}(x,y)$ is defined in (\ref{eq:FxyGGE}).  Since we are only interested in the two-giant graviton contribution, we can focus on the cases when the $k_i$'s and $\tilde k_i$'s are all one.  Then
\begin{align}
    G_N^{(p=0;1,1)}(x,y)&=x^{2N}Z_2(x^{-1},xy),\nn\\
    G_N^{(p=0;\tilde1,\tilde1)}(x,y)&=y^{2N}Z_2(xy,y^{-1}),\nn\\
    G_N^{(p=0;1,\tilde1)}(x,y)&=(xy)^NxyZ_1(x^{-1},xy)Z_1(xy,y^{-1}).
\end{align}
Note that this has the same structure as the D3-brane indices, (\ref{eq:hatZ2s}), except that $Z_2(\cdot,\cdot)$ is to be computed by contour integration.  Since $Z_1(\cdot,\cdot)$ is unambiguous, we do find $G_N^{(p=0;1,\tilde 1)}(x,y)=(xy)^N\hat Z_{(1,1)}(x,y)$ as expected.

The remaining term to consider is the $p=1$ term, $G_N^{(\ell,\tilde\ell)}$, which can be obtained from (\ref{eq:GMStot}) and (\ref{eq:pairs})
\begin{align}
    G_N^{(\ell,\tilde\ell)}(x,y)&=(-1)^{\ell+\tilde\ell+1}(x^\ell y^{\tilde\ell})^N(xy)^{\ell\tilde\ell}x^{\ell(\ell-1)/2}y^{\tilde\ell(\tilde\ell-1)/2}\nn\\
    &\quad\times\int\fft{dz}{2\pi iz}z^{N+\ell+\tilde\ell-1}
    \fft{(y;y)_\infty^2}{(zx^\ell y;y)_\infty(z^{-1}x^{-\ell};y)_\infty}
    \fft{(x;x)_\infty^2}{(zy^{\tilde\ell}x;x)_\infty(z^{-1}y^{-\tilde\ell};x)_\infty}.
\end{align}
Note that this expression originally contained contour integrals over $z_1$ and $z_2$, but we have eliminated one of these integrals by setting $z_2=1$ and relabeling $z_1\to z$.  In order to count the number of giant gravitons contained in this expression, we note the $N$ dependence of $G_N^{(\ell,\tilde\ell)}$ arises from a combination of $(x^\ell y^{\tilde\ell})^N$ and $z^N$, where $z$ will be replaced by its values at the poles.  Since $z$ is integrated along the unit circle, the possible poles inside the unit circle are of the form
\begin{equation}
    z=\fft{y^k}{x^\ell},\qquad z=\fft{x^k}{y^{\tilde\ell}},\qquad \{\ell,\tilde\ell\}\ge1,\quad k\ge0.
\end{equation}
The $N$-dependence of $G_N^{(\ell,\tilde\ell)}$ is then of the form $(y^{k+\tilde\ell})^N$ or $(x^{k+\ell})^N$, for the first or second type poles, respectively.

Suppose we take the first type pole, $z=y^k/x^\ell$.  In order for this to contribute to the two-giant graviton index, we can have either $k=0$, $\tilde\ell=2$ or $k=1$, $\tilde\ell=1$, so that $(y^{k+\tilde\ell})^N=y^{2N}$.  For $k=0$, the pole is at $z=1/x^\ell$, which is always outside the unit circle, since we are assuming an expansion around small fugacities, $|x|<1$.  For $k=1$, the pole is at $z=y/x^\ell$.  However, for $\ell>1$, this pole lies outside the unit circle, assuming symmetric scaling of the fugacities, $|x|\sim|y|<1$.  Hence the only possible pole that may contribute is the one at $z=y/x$, corresponding to $k=\ell=\tilde\ell=1$, and moreover this only contributes if $|y/x|<1$.  A similar argument applies to the second type pole, $z=x/y$, which contributes for $|x/y|<1$.  We then find
\begin{equation}
    G_N^{(1,\tilde 1)}(x,y)=\begin{cases}-x^{2N+1}Z_1(x^{-2},x^2y),&|x/y|<1;\\-y^{2N+1}Z_1(y^2x,y^{-2}),&|y/x|<1,\end{cases}
\end{equation}
where $Z_1(\cdot,\cdot)$ is the $U(1)$ Schur index, with the explicit form given in (\ref{eq:zeeone}).  Note that this one pair contribution to Murthy's $G_N^{(2)}$ has the exact same form as the two-giant contribution within $G_N^{(1)}$, shown in (\ref{eq:GN1to2gg}).

We now have all the components we need to compare with the D3-brane giant graviton index.  This is summarized in Table~\ref{tbl:twogg}.  While $\hat Z_{(1,1)}$ is reproduced exactly from the expansion of Murthy's $G_N^{(2)}$, the situation is more complicated for $\hat Z_{(2,0)}$ and $\hat Z_{(0,2)}$.  Since these expressions are mapped into each other by $x\leftrightarrow y$, we can focus just on $\hat Z_{(2,0)}$.  As discussed in \cite{Arai:2020qaj,Beccaria:2023zjw}, the D3-brane index is simply
\begin{equation}
    \hat Z_{(2,0)}(x,y)=Z_2^{(\mathrm{ac})}(x^{-1},xy)\qquad(\mbox{by analytic continuation)}.
\end{equation}
However, Murthy's expansion yields
\begin{equation}
    \hat Z_{(2,0)}(x,y)=Z_2^{(\mathrm{ci})}(x^{-1},xy)-(1+\theta(|y|-|x|)) xZ_1(x^{-2},x^2y)\qquad(\mbox{by contour integration)},
\label{eq:Z20ci}
\end{equation}
where $\theta(x)=1$ if $x>0$ and $0$ otherwise is the unit step function.

\begin{table}[t]
\begin{tabular}{l|l|ll}
&\hfil$G_N^{(1)}$&\multicolumn{2}{c}{$G_N^{(2)}$}\\
&&\hfil$p=0$&\hfil$p=1$\\\hline
$\hat Z_{(2,0)}$&$-xZ_1(x^{-2},x^2y)$&$Z_2(x^{-1},xy)$&$-xZ_1(x^{-2},x^2y)$ if $|x/y|<1$\\
$\hat Z_{(1,1)}$&&$xyZ_1(x^{-1},xy)Z_1(xy,y^{-1})$\\
$\hat Z_{(0,2)}$&$-yZ_1(y^2x,y^{-2})$&$Z_2(xy,y^{-1})$&$-yZ_1(y^2x,y^{-2})$ if $|y/x|<1$
\end{tabular}
\caption{Contributions to the two D3-brane giant graviton index from Murthy's $G_N^{(1)}$ and $G_N^{(2)}$.}
\label{tbl:twogg}
\end{table}

Since the D3-brane index is well-defined and independent of the magnitude of $|x/y|$, it seems unusual that the expression, (\ref{eq:Z20ci}), has such a dependence.  The resolution of this puzzle is that the contour integral expression for $Z_2(x^{-1},xy)$ must necessarily also depend on the magnitude of $|x/y|$ such that the dependence ends up dropping out in the final expression.  To see how this works in practice, we start with $Z_2(x,y)$ given from (\ref{eq:N=4Schur}), and write
\begin{align}
    Z_2(x^{-1},xy)&=\fft12\fft{(y;y)_\infty^4}{(x^{-1};y)_\infty^2(xy;y)_\infty^2}\nn\\
    &\quad\times\int\fft{dz}{2\pi iz}(1-z)(1-z^{-1})\fft{(zy;y)_\infty^2(z^{-1}y;y)_\infty^2}{(zx^{-1};y)_\infty(z^{-1}x^{-1};y)_\infty(zxy;y)_\infty(z^{-1}xy;y)_\infty},
\label{eq:Z2x-1ci}
\end{align}
where $z$ is integrated on the unit circle.  Note that we have removed one of the two contour integrals by setting $z_2=1$ and $z_1=z$.

\begin{table}[t]
\centering
\begin{tabular}{l|c|c}
pole ($k=0,1,2,\ldots$)&analytic continuation&contour integration\\\hline
$0$&\checkmark&\checkmark\\
$x$&&\checkmark\\
$x^{-1}$&\checkmark&\\
$x/y$&&\checkmark~ if $|x/y|<1$\\
$y/x$&\checkmark&\checkmark~ if $|y/x|<1$\\
$xy^{k+1}$&\checkmark&\checkmark\\
$x/y^{k+2}$&&\\
$x^{-1}y^{k+2}$&\checkmark&\checkmark\\
$x^{-1}/y^{k+1}$&&
\end{tabular}
\caption{Poles in the integrand of $Z_2(x^{-1},xy)$ given in (\ref{eq:Z2x-1ci}).  The pole at $z=0$ is a double pole, while the remaining poles are single poles.  The checkmarks indicate that the pole is inside the contour for the case of analytic continuation from $Z_2(x,y)$ or for the case of direct contour integration of $Z_2(x^{-1},xy)$.}
\label{tbl:poles}
\end{table}

As written, the above expression has a double pole at $z=0$ and single poles at $z=xy^{-k}$, $x^{-1}y^k$, $x^{-1}y^{-k-1}$ and $xy^{k+1}$ for $k=0,1,2,\ldots$.  The evaluation of this integral depends on which poles are chosen to lie within the integration contour.  The analytic continuation expression for $Z_2(x^{-1},xy)$ corresponds to assuming both arguments $x^{-1}$ and $xy$ lie within the contour, which is equivalent to the pole selection criteria outlined in \cite{Imamura:2021ytr}.  In this case, the poles contained within the unit circle are summarized in Table~\ref{tbl:poles}.  The result of analytic continuation was obtained in \cite{Arai:2020qaj,Beccaria:2023zjw}, and takes the form
\begin{equation}
    Z_2^{(\mathrm{ac})}(x^{-1},xy)=-\fft{x^7(x-2y)}{(x-y)^2y(x+y)}+\fft{x^5}{y^2}(2y^2-x^2)+\fft{x^3}{y^3}(-x^6+2x^3y^3+2y^6)+\cdots.
\end{equation}
Here the superscript (ac) is a reminder that $Z_2(\cdot,\cdot)$ defined in (\ref{eq:N=4Schur}) is to be evaluated by analytic continuation.

On the other hand, Murthy's expression, (\ref{eq:Z20ci}), involves a direct evaluation of the contour integral in (\ref{eq:Z2x-1ci}) for $|x|,|y|<1$.  In this case, we encircle the double pole at the origin along with single poles at $x$ as well as $x^{-1}y^{k+2}$ and $xy^{k+1}$ for $k=0,1,2,\ldots$ and either the single pole at $x/y$ or at $y/x$ depending on the magnitude of $|x/y|$, as summarized in Table~\ref{tbl:poles}.  Comparing the analytic continuation and contour integration cases, we see that the difference lies only in the treatment of the poles at $x$, $x^{-1}$, $x/y$ and $y/x$.  In particular, we find the relation
\begin{equation}
    Z_2^{(\mathrm{ci})}(x^{-1},xy)=Z_2^{(\mathrm{ac})}(x^{-1},xy)+\mbox{Res}\big|_x-\mbox{Res}\big|_{x^{-1}}+\theta(|y|-|x|)(\mbox{Res}\big|_{x/y}-\mbox{Res}\big|_{y/x}).
\end{equation}
The residues are easy to evaluate from (\ref{eq:Z2x-1ci}), and we find
\begin{equation}
    \mbox{Res}\big|_x=-\mbox{Res}\big|_{x^{-1}}=\mbox{Res}\big|_{x/y}=-\mbox{Res}\big|_{y/x}=\fft12xZ_1(x^{-2},x^2y),
\end{equation}
where there is no ambiguity in the definition of $Z_1(\cdot,\cdot)$ given in (\ref{eq:zeeone}).  As a result, we find the identity
\begin{equation}
    Z_2^{(\mathrm{ci})}(x^{-1},xy)=Z_2^{(\mathrm{ac})}(x^{-1},xy)+(1+\theta(|y|-|x|))xZ_1(x^{-2},x^2y).
\end{equation}
Inserting this into Murthy's expression, (\ref{eq:Z20ci}), we find that the unwanted terms indeed cancel, and we are left with 
$\hat Z_{(2,0)}(x,y)=Z_2^{(\mathrm{ac})}(x^{-1},xy)$, which verifies that Murthy's expression for two giant gravitons indeed matches the D3-brane expansion of the index.

We expect that this analysis can be extended to three or more giant gravitons.  The index for $k$ D3-brane giants will receive contributions from Murthy's $G_N^{(m)}$ with $m\le k$, where $G_N^{(m)}$ itself admits an expansion over a set of $w$ poles as given in (\ref{eq:GMmsum}).  This sum includes more and more terms as we consider a larger number of giant gravitons.  However, as we have seen, there is a subset of terms in the zero-pair sector given by $G_N^{(p=0;\mathbf 1,\mathbf 1)}(x,y)$ in (\ref{eq:murthytoafim}) that superficially matches the D3-brane index.  Organizing Murthy's expansion in powers of $x^N$ and $y^N$, we then have
\begin{equation}
    \hat Z_{(n_1,n_2)}(x,y)=(x^{n_1}y^{n_2})^NZ_{n_1}^{(\mathrm{ci})}(x^{-1},xy)(xy)^{n_1n_2}Z_{n_2}^{(\mathrm{ci})}(xy,y^{-1})+\cdots,
\end{equation}
where the additional terms arise from $p>0$ contributions in (\ref{eq:GMStot}) as well as from non-minimal terms in $G_N^{(m)}$ with $m<n_1+n_2$.  In general, there can be a large number of such terms.  However, they must necessarily contribute in a precise manner to shift from the contour integration expression to the analytic continuation expression for the index \cite{Arai:2020qaj,Beccaria:2023zjw}
\begin{equation}
    \hat Z_{(n_1,n_2)}(x,y)=(x^{n_1}y^{n_2})^NZ_{n_1}^{(\mathrm{ac})}(x^{-1},xy)(xy)^{n_1n_2}Z_{n_2}^{(\mathrm{ac})}(xy,y^{-1}).
\end{equation}
It would be good to see how this occurs in practice.

\section{The 1/16 BPS index}
\label{sec:1/16}

We now conclude with a brief examination of the 1/16-BPS index, which we take to be a function of fugacities $p$, $q$ and $y_i$ under the constraint $pq=y_1y_2y_3$.  In particular, our goal is to identify components of Murthy's $G_N ^{(m)}(p,q;y_i)$ with their counterparts in the D3-brane indices $\hat Z_{(n_1, n_2, n_3)}(p,q;y_i)$.  

Before turning to the giant graviton expression, recall that the 1/16 BPS index can be obtained from the single letter index $f(p,q;y_i)$ given in (\ref{eq:f1/16}).  The resulting matrix integral can be written in terms of an elliptic hypergeometric integral \cite{Dolan:2008qi,Spiridonov:2010qv}
\begin{equation}
Z_N(p,q;y_a)=\fft{((p;p)_\infty(q;q)_\infty)^{N}}{N!}\prod_{a=1}^3\Gamma(y_a;p,q)^{N}\oint\prod_{i=1}^{N}\fft{dz_i}{2\pi iz_i}\prod_{i\ne j}^N\fft{\prod_{a=1}^3\Gamma(\fft{z_i}{z_j}y_a;p,q)}{\Gamma(\fft{z_i}{z_j};p,q)},
\label{eq:EHI}
\end{equation}
where the elliptic gamma function is defined by
\begin{equation}
    \Gamma(z;p,q)=\prod_{m,n\ge0}\fft{1-p^{m+1}q^{n+1}/z}{1-p^mq^n}.
\end{equation}

\subsection{Murthy's giant graviton expansion}

In the 1/16 BPS case, the single letter index $f(p,q;y_i)$ yields the effective single-particle function
\begin{equation}
    \begin{split}
        \hat{f}(p,q;y_i)&= \frac{(1-p)(1-q)}{(1-y_1) (1-y_2)(1-y_3)} - 1 \\ 
        &=\sum_{\vec{n}}{\strut}^\prime \vec{y}^{\ \vec{n}} + \sum_{\vec{n}} \vec{y} ^{\ \vec{n}+1} - p \vec{y} ^{\ \vec{n}} - (p \leftrightarrow q) ~,
     \end{split}
     \label{eq:1/16fhat}
\end{equation}
where $\vec{n}$ denotes all nonnegative $(n_1,n_2,n_3)$, the prime excludes $(0,0,0)$, and $\vec{y} ^{\ \vec{n}}$ is shorthand for $y_1 ^{n_1} y_2 ^{n_2} y_3 ^{n_3}$.  Inserting $\hat f(p,q;y_i)$ into \eqref{eq:EnGG2} and working out the Plethystic exponential gives the integral expression
\begin{align}
    G_N ^{(m)}(p,q;\vec y\,) &= \frac{(-1)^m}{m! ^2} \int\prod_i\left( \frac{dz_i}{2\pi i z_i} \frac{dw_i}{2\pi i w_i}\right)\left(\frac{w_1\cdots w_m}{z_1\cdots z_m}\right)^{N+m}\frac{\prod_{i\neq j} (1-\tfrac{z_i}{z_j})(1-\tfrac{w_i}{w_j})}{\prod_{i,j=1} ^m (1-\tfrac{w_j}{z_i})^2}\nn \\
    &\quad \times \prod _{\vec{n}}\strut' \frac{(1-\vec{y}^{\ \vec{n}})^{2m} \prod_{i\neq j} (1-\tfrac{z_i}{z_j} \vec{y} ^{\ \vec{n}})(1-\tfrac{w_i}{w_j}\vec{y} ^{ \ \vec{n}})}{\prod_{i,j} (1-\tfrac{z_i}{w_j} \vec{y}^{\ \vec{n}})(1-\tfrac{w_i}{z_j} \vec{y} ^{\ \vec{n}} )}\nn\\
    &\quad\times\prod _{\vec{n}} \frac{(1-\vec{y}^{\ \vec{n}+1})^{2m} \prod_{i\neq j} (1-\tfrac{z_i}{z_j} \vec{y} ^{\ \vec{n}+1})(1-\tfrac{w_i}{w_j}\vec{y} ^{ \ \vec{n}+1})}{\prod_{i,j} (1-\tfrac{z_i}{w_j} \vec{y}^{\ \vec{n}+1})(1-\tfrac{w_i}{z_j} \vec{y} ^{\ \vec{n}+1} )}\nn \\ 
    &\quad\times \prod _{\vec{n}} \frac{\prod_{i, j} (1-\tfrac{z_i}{w_j} p \vec{y} ^{\ \vec{n}})(1-\tfrac{w_i}{z_j} p \vec{y} ^{\ \vec{n}})}{(1-p\vec{y} ^{\ \vec{n}})^{2m} \prod_{i\neq j} (1-\tfrac{z_i}{z_j} p\vec{y} ^{\ \vec{n}})(1-\tfrac{w_i}{w_j} p\vec{y} ^{\ \vec{n}})} \cdot (p \leftrightarrow q) ~. 
\label{eq:GNmwz}
\end{align}
As we have seen with the Schur index, we may first perform the $w_i$ integrals and in this way split $G_N ^{(m)}(p,q;\vec y\,)$ into a sum of terms, all with different sets of poles and their residues.  While we will not be completely systematic in our treatment, we may highlight some important contributions to the giant graviton expansion.

Our first goal is to highlight only the giant gravitons moving on a single rotation plane.  These contribute to $\hat Z_{(m,0,0)}(p,q;\vec y\,)$ and its permutations, and are obtained by isolating the contributions to $G_N ^{(m)}(p,q;\vec y\,)$ with $N$-dependence of the form $y_i ^N$. We can pick $y_1 ^N$ without loss of generality, which requires the choice of $w_i$ poles:
\begin{equation}
    w_i = y_1 z_i\quad \text{ for }\quad i=1,\ldots,m     ~,
\label{eq:wiziy1 pole}
\end{equation}
up to permutations.  For this pole choice, \eqref{eq:GNmwz} reduces to
\begin{equation}
    \begin{split}
        G_N ^{(m)}(p,q;\vec y\,) \Big\vert_{w_i = y_1 z_i} &= \frac{(-)^m}{m! ^2} y_1 ^{mN + m(m-1)} \int \prod_i \frac{dz_i}{2\pi i z_i}  \frac{\prod_{i\neq j} (1-\tfrac{z_i}{z_j} )^2}{(1-y_1)^{2m} \prod_{i\neq j}(1-y_1 \tfrac{z_i}{z_j})^2 } \\
        & y_1 ^m  \left(\prod_{\vec{n}}\frac{(1-\vec{y} ^{\ \vec{n}'})^2}{(1-y_1 \vec{y} ^{\ \vec{n}'})(1-y_1 ^{-1} \vec{y} ^{\ \vec{n}'})}\right)^m \left(\prod_{\vec{n}}\frac{(1-\vec{y} ^{\ \vec{n}+1})^2}{(1-y_1 \vec{y} ^{\ \vec{n}+1})(1-y_1 ^{-1} \vec{y} ^{\ \vec{n}+1})}\right)^m  \\
        & \left(\prod_{\vec{n}}\frac{(1-p\vec{y} ^{\ \vec{n}})^2}{(1-y_1 p\vec{y} ^{\ \vec{n}})(1-y_1 ^{-1} p\vec{y} ^{\ \vec{n}})}\right)^{-m} \cdot (p \leftrightarrow q)\\
        & \prod _{\vec{n},i\neq j} \frac{ (1-\tfrac{z_i}{z_j}  \vec{y} ^{\ \vec{n}'})^2}{ (1-\tfrac{z_i}{z_j}   y_1 \vec{y}^{\ \vec{n}'})(1-\tfrac{z_i}{z_j}  y_1 ^{-1} \vec{y} ^{\ \vec{n}'} )}  \frac{ (1-\tfrac{z_i}{z_j}  \vec{y} ^{\ \vec{n}+1})^2 }{ (1-\tfrac{z_i}{z_j}   y_1 \vec{y}^{\ \vec{n}+1})(1-\tfrac{z_i}{z_j}  y_1 ^{-1} \vec{y} ^{\ \vec{n}+1} )}  \\ 
        & \prod _{\vec{n},i\neq j} \frac{(1-\tfrac{z_i}{z_j}   y_1 p\vec{y}^{\ \vec{n}})(1-\tfrac{z_i}{z_j}  y_1 ^{-1} p\vec{y} ^{\ \vec{n}} )}{ (1-\tfrac{z_i}{z_j}  p\vec{y} ^{\ \vec{n}})^2} \cdot (p \leftrightarrow q) ~. 
    \end{split}
    \label{eq:GNmy1zeta}
\end{equation}
There are partial cancellations of note in the product terms with overall exponent $\pm m$:
\begin{align}
    \prod_{\vec{n}}\frac{(1-\vec{y} ^{\ \vec{n}'})^2}{(1-y_1 \vec{y} ^{\ \vec{n}'})(1-y_1 ^{-1} \vec{y} ^{\ \vec{n}'})} &=  (1-y_1 ^{-1})(1-y_1) \sideset{}{'}\prod_{n_2,n_3\geq 0}  \frac{(1-y_2 ^{n_2} y_3 ^{n_3})}{(1-y_1 ^{-1} y_2 ^{n_2} y_3 ^{n_3})} ~,\nn \\ 
    \prod_{\vec{n}}\frac{(1-\vec{y} ^{\ \vec{n}+1})^2}{(1-y_1 \vec{y} ^{\ \vec{n}+1})(1-y_1 ^{-1} \vec{y} ^{\ \vec{n}+1})} &= \sideset{}{}\prod_{n_2,n_3\geq 0} \frac{(1-y_1 y_2 ^{n_2 +1} y_3 ^{n_3 +1})}{(1- y_2 ^{n_2+1} y_3 ^{n_3+1})}~,\nn\\ 
    \prod_{\vec{n}}\frac{(1-p\vec{y} ^{\ \vec{n}})^2}{(1-y_1 p\vec{y} ^{\ \vec{n}})(1-y_1 ^{-1} p\vec{y} ^{\ \vec{n}})} &= \prod_{n_2,n_3\geq 0} \frac{(1- py_2 ^{n_2} y_3 ^{n_3 })}{(1-y_1 ^{-1} p y_2 ^{n_2} y_3 ^{n_3})}~. 
\end{align}
Moreover, the products involving $z_i/z_j$ undergo identical partial cancellations.  As a result, we find
\begin{equation}
\label{eq: 1/16 m giants}
    \begin{split}
        G_N ^{(m)}(p,q;\vec y\,) \Big\vert_{w_i = y_1 z_i}&= \frac{y_1 ^{mN}}{m!^2}  \left[(y_2 ; y_2)_\infty (y_3 ; y_3)_\infty \right]^m \Gamma\left(p,q,y_1^{-1};y_2, y_3\right) ^m \\
        &\quad\times\int \prod_i \frac{dz_i}{2\pi i z_i} \prod_{i\neq j} \fft{\Gamma\left(\frac{z_i}{z_j} p,\frac{z_i}{z_j} q,\frac{z_i}{z_j} \frac{1}{y_1}; y_2,y_3\right)}{\Gamma\left(\frac{z_i}{z_j} ; y_2,y_3\right)},
    \end{split}
\end{equation}
where we have introduced the shorthand notation
\begin{equation}
        \Gamma(z_1, \ldots, z_n; p, q) = \Gamma(z_1; p, q) \cdots \Gamma(z_n; p, q).
\end{equation}
Comparing this expression with (\ref{eq:EHI}) and taking into account the $m!$ distinct possible choices of poles $w_i = y_1 z_j$, we find
\begin{equation}
    G_N ^{(m)}(p,q;\vec y\,)\Big\vert_{w_i = y_1 z_j} = y_1 ^{mN}Z_m^{(\mathrm{ci})}(y_2,y_3;y_1^{-1},p,q)=y_1^{mN}\hat Z_{(m,0,0)}^{(\mathrm{c i})}(p,q;\vec y\,).
\end{equation}
In particular, this mapping of fugacities agrees with (\ref{eq:fvamap}) identified previously for the 1/16 BPS giant graviton expansion \cite{Imamura:2021ytr,Gaiotto:2021xce,Lee:2022vig}.

As in the case of the Schur index, this identification of $ G_N ^{(m)}(p,q;\vec y\,)\vert_{w_i = y_1 z_j}$ with the wrapped D3-brane index $\hat Z_{(m,0,0)}(p,q;\vec y\,)$ is incomplete in itself, as the latter is given by analytic continuation and not by contour integration with standard pole selection.  Nevertheless, we can see hints at how the full connection between $G_N^{(m')}(p,q;\vec y\,)$ with $m'\le m$ and $\hat Z_{(m,0,0)}(p,q;\vec y\,)$ will develop.

We now consider the main contribution to $\hat Z_{(m_1,m_2,0)}(p,q;\vec y\,)$.  For this, we are interested in the leading behavior in $G_N^{(m)}(p,q;\vec y\,)$ that goes as a product $y_i ^{m_i N} y_j ^{m_j N}$. We thus consider the following $w_i$ poles:
\begin{equation}
\label{eq: 1/16 wz m1 m2 poles}
\begin{split}
    w_a &= y_1 z_a  \ , \ a=1, \ldots , m_1 \\ 
    w_b &= y_2 z_b \ , \ b=m_1 + 1 , \ldots , m_1+m_2 ~,
\end{split}
\end{equation}
where without loss of generality we choose $y_1$ and $y_2$. The form for general $y_i$ and $y_j$ can be obtained with suitable cyclical permutation.  
Additionally, just like for the Schur index, we obtain a combinatorial factor counting for the ways in which $m$ can be split into $m_1$ and $m_2$, combining with the $1/(m!)^2$ prefactor to give a final $\tfrac{1}{m_1 !} \tfrac{1}{m_2 !}$ prefactor. 

Applying the pole choice \eqref{eq: 1/16 wz m1 m2 poles} to the expression \eqref{eq:GNmwz} leads to an involved integrand, which upon further inspection, simplification and partial cancellation of the type previously encountered for the single $y_i$ pole case, separates multiplicatively into:
\begin{equation}
\label{eq: 1/16 Gm m1 m2}
    \begin{split}
        G_N ^{(m)}(p,q;\vec y\,)  \Big\vert_{w_a = y_1 z_a, w_b = y_2 z_b}&=
        \frac{y_1 ^{m_1 N} y_2 ^{m_2 N}}{m_1 ! m_2 !} \left[\Gamma(p,q,y_1 ^{-1}; y_2 , y_3) (y_2 ; y_2)_\infty (y_3 ; y_3)_\infty \right]^{m_1}\\
        &\kern-6em\times\left[\Gamma(p,q,y_2 ^{-1}; y_1 , y_3) (y_1 ; y_1)_\infty (y_3 ; y_3)_\infty \right]^{m_2} \\ 
        &\kern-6em\times \int \prod_{a=1} ^{m_1} \frac{dz_a}{2\pi i z_a} \prod_{b=m_1 +1} ^{m_1+m_2} \frac{dz_b}{2\pi i z_b} \prod_{a_1 \neq a_2} \fft{\Gamma\left(\frac{z_{a_1}}{z_{a_2}} p, \frac{z_{a_1}}{z_{a_2}} q, \frac{z_{a_1}}{z_{a_2}} y_1 ^{-1}; y_2,y_3 \right)}{\Gamma\left(\frac{z_{a_1}}{z_{a_2}}; y_2, y_3 \right)}  \\
        &\kern6.5em \prod_{b_1 \neq b_2}\fft{\Gamma\left(\frac{z_{b_1}}{z_{b_2}} p, \frac{z_{b_1}}{z_{b_2}} q, \frac{z_{b_1}}{z_{b_2}} y_2 ^{-1}; y_1,y_3 \right)}{\Gamma\left(\frac{z_{b_1}}{z_{b_2}}; y_1, y_3 \right)}\\ 
        &\kern-5em\prod_{a=1} ^{m_1} \prod_{b=m_1 +1} ^{m_1 + m_2} \fft{\left(\frac{z_a}{z_b} py_2 ^{-1} ; y_3  \right)_\infty \left(\frac{z_a}{z_b} q y_2 ^{-1} ; y_3  \right)_\infty }{\left(\frac{z_a}{z_b} y_2 ^{-1} ; y_3  \right)_\infty\left( \frac{z_a}{z_b} y_1 y_3  ; y_3  \right)_\infty} \fft{\left(\frac{z_b}{z_a} py_1 ^{-1}; y_3  \right)_\infty\left( \frac{z_b}{z_a} q y_1 ^{-1} ; y_3  \right)_\infty }{\left(\frac{z_b}{z_a} y_1 ^{-1} ; y_3  \right)_\infty\left(\frac{z_b}{z_a} y_2 y_3  ; y_3  \right)_\infty }~. 
    \end{split}
\end{equation}
This has a natural interpretation from the D3-brane point of view.  The first two lines inside the integral correspond to the vector multiplet contributions with single letter indices $f_v^{(1)}(p,q;\vec y\,)$ and $f_v^{(2)}(p,q;\vec y\,)$ of (\ref{eq:1/16singleletter}), while the last line corresponds to the bifundamental hypermultiplet with single letter index $f_h^{(1,2)}(p,q;\vec y\,)$ of (\ref{eq:1/16singleletter}).  Thus we have the match
\begin{equation}
    G_N ^{(m)} (p,q;\vec y\,)\Big\vert_{w_a = y_1 z_a, w_b = y_2 z_b} = y_1 ^{m_1 N} y_2 ^{m_2 N} \hat Z_{(m_1, m_2, 0)}^{(\mathrm{ci})}(p,q;\vec y\,),
\end{equation}
with (ci) reminding us that the contour integration is to be performed with the standard choice of poles.  We, of course, expect additional contributions to arise that will shift $\hat Z_{(m_1, m_2, 0)}^{(\mathrm{ci})}(p,q;\vec y\,)$ to the final form of the giant graviton index (corresponding to the analytic continuation result).

\subsection{The Schur limit}
The Schur limit of the aforementioned expressions for the 1/16 BPS index giant graviton expansion is an appropriate check, that brings into connection the results from Sec.~\ref{sec:murthyschur}. This limit takes the form of identifying $q$ with one of the $y_i$, and at the level of the single-particle partition function \eqref{eq:1/16fhat}:
\begin{equation}
    q= y_i \quad\implies\quad \hat{f}(p,q;\vec y\,) = \frac{(1-p)}{(1-y_j)(1-y_k)} - 1 ~,
\end{equation}
where $y_j$ and $y_k$ are the two other $y$-fugacities in addition to the choice of $y_i$. Making use of $pq=y_1 y_2 y_3$, we note that $p=y_j y_k$, and thus we can re-express the single-letter index in terms of just $y_j$ and $y_k$, to be relabeled as $x$ and $y$ respectively:
\begin{equation}
    \hat{f}(x,y) = \frac{1-xy}{(1-x)(1-y)} - 1 ~. 
\end{equation}
Our goal is to connect the giant gravitons expressions, \eqref{eq: 1/16 m giants} and \eqref{eq: 1/16 Gm m1 m2} in the Schur limit to the corresponding 1/8 BPS expression \eqref{eq:murthytoafim}.

We start with the single rotation plane giant graviton expression, \eqref{eq: 1/16 m giants}, as well as its $\{y_1,y_2,y_3\}$ permutations.  To be concrete, we take the Schur limit by setting $q=y_1$ (and thus $p=y_2 y_3$).  The three cases to consider are then the stack of $w_i=y_1z_i$ poles, the stack of $w_i = y_2 z_i$ poles and the stack of $w_i = y_3 z_i$ poles.  First, we immediately note that \eqref{eq: 1/16 m giants}, corresponding to $w_i=y_1z_i$, features the term $\Gamma(p; y_2, y_3) = \Gamma(y_2 y_3 ; y_2 , y_3)$, once we set $q=y_1$.  However, upon further inspection:
\begin{equation}
    \Gamma(p=y_2y_3; y_2, y_3) =\prod_{j,k=0}^\infty\fft{1-y_2 ^{j}y_3 ^{k}}{1-y_2 ^{j+1} y_3 ^{k+1} } = 0 ~. 
\end{equation}
This vanishes identically because of a zero in the numerator.  There is a similar ${z_i}/{z_j}$-dependent term in the integrand of \eqref{eq: 1/16 m giants} that also vanishes in this way. We thus find that:
\begin{equation}
    G_N ^{(m)}(p,q;\vec y\,) \Big\vert_{w_i = y_1 z_i, q=y_1} =0 ~. 
\end{equation}

While this vanishes identically, the $w_i = y_2 z_i$ pole choice under the $q=y_1$ setting evaluates to a non-trivial result. Relabeling $y_2 = x$, $y_3 = y$, we simplify the following product in the analog to \eqref{eq: 1/16 m giants}: 
\begin{equation}
    \left[(y_1 ; y_1)_\infty (y_3 ; y_3)_\infty \right]^m \Gamma\left(p,q,y_2^{-1};y_1, y_3\right) ^m = \frac{(y;y)^{2m} _\infty }{(x^{-1} y)_\infty ^m (xy; y)_\infty ^m }~. 
\end{equation}
Any terms in the product that involved $y_1 = q$ cancel away, and the elliptic gamma fractions simplify as well, leaving only $x$ and $y$ $q$-Pochhammer terms. The integrand in \eqref{eq: 1/16 m giants} simplifies similarly leading to: 
\begin{equation}
    \prod_{i\neq j}\fft{ \Gamma\left(\frac{z_i}{z_j} p,\frac{z_i}{z_j} q,\frac{z_i}{z_j}y_2^{-1}; y_1,y_3\right)}{\Gamma\left(\frac{z_i}{z_j} ; y_1,y_3\right)} = \prod_{i\neq j} \frac{(\tfrac{z_i}{z_j} ; y)_\infty (y\tfrac{z_i}{z_j}; y)_\infty}{(x^{-1} \tfrac{z_i}{z_j}; y)_\infty (xy \tfrac{z_i}{z_j}; y)_\infty } ~. 
\end{equation}
We thus find that the $q=y_1$ limit gives:
\begin{equation}
    G_N ^{(m)}(p,q;\vec y\,)\Big \vert_{w_i = y_2 z_i, q=y_1} = y_2 ^{mN} Z_m^{(\mathrm{ci})} (y_2 ^{-1} , y_2 y_2) ~,
\end{equation}
where $Z_N(x,y)$ is the integral expression for the Schur index, (\ref{eq:N=4Schur}).  Since $y_2$ and $y_3$ play interchangeable roles in \eqref{eq: 1/16 m giants}, we immediately find that the $w_i = y_3 z_i$ pole choice gives
\begin{equation}
     G_N ^{(m)}(p,q;\vec y\,)\Big  \vert_{w_i = y_3 z_i, q=y_1} = y_3 ^{mN} Z_m^{(\mathrm{ci})} (y_3 ^{-1} , y_2 y_3) ~. 
\end{equation}
This analysis can be summarized as:
\begin{equation}
q=y_1 : \ G_N ^{(m)}(p,q,\vec y\,) =
\begin{cases} 
      &w_i = y_1 z_i: \ \hat{Z}_{(m,0,0)}^{(\mathrm{ci})}(y_2y_3,y_1;\vec y\,) = 0 ~, \\
      &w_i = y_2 z_i: \ \hat{Z}_{(0,m,0)}^{(\mathrm{ci})}(y_2y_3,y_1;\vec y\,) = Z_m^{(\mathrm{ci})} (y_2 ^{-1}, y_2 y_3) ~, \\
      &w_i = y_3 z_i: \ \hat{Z}_{(0,0,m)}^{(\mathrm{ci})}(y_2y_3,y_1;\vec y\,) = Z_m^{(\mathrm{ci})} (y_3 ^{-1}, y_2 y_3) ~.    
\end{cases}
\end{equation}
Since we have taken the $q=y_1$ Schur limit, giant gravitons moving along the first rotation plane decouple, and we are left with Schur index contributions only from $\hat{Z}_{(0,m,0)}$ and $\hat{Z}_{(0,0,m)} $.

One further consistency check with the Schur index expression, \eqref{eq:murthytoafim}, involves the more complicated pole choice \eqref{eq: 1/16 wz m1 m2 poles}. The 1/16 BPS expression, in this case \eqref{eq: 1/16 Gm m1 m2}, can be decomposed into two factorizable prefactors and integrands that separate into $m_1$ and $m_2$ terms:
\begin{equation}
    \begin{split}
        &\frac{y_1 ^{m_1 N}}{m_1 !} \left[\Gamma(p,q,y_1 ^{-1}; y_2 , y_3) (y_2 ; y_2)_\infty (y_3 ; y_3)_\infty \right]^{m_1} \int \prod_{a=1} ^{m_1} \frac{dz_a}{2\pi i z_a} \prod_{a_1 \neq a_2} \fft{\Gamma\left(\frac{z_{a_1}}{z_{a_2}} p, \frac{z_{a_1}}{z_{a_2}} q, \frac{z_{a_1}}{z_{a_2}} y_1 ^{-1}; y_2,y_3 \right) }{ 
     \Gamma\left(\frac{z_{a_1}}{z_{a_2}}; y_2, y_3 \right)} , \\ 
    &\frac{y_2 ^{m_2 N}}{m_2 !} \left[\Gamma(p,q,y_2 ^{-1}; y_1 , y_3) (y_1 ; y_1)_\infty (y_3 ; y_3)_\infty \right]^{m_2} \int \prod_{b=m_1 +1} ^{m_1+m_2} \frac{dz_b}{2\pi i z_b} \prod_{b_1 \neq b_2} \fft{\Gamma\left(\frac{z_{b_1}}{z_{b_2}} p, \frac{z_{b_1}}{z_{b_2}} q, \frac{z_{b_1}}{z_{b_2}} y_2 ^{-1}; y_1,y_3 \right)}{ \Gamma\left(\frac{z_{b_1}}{z_{b_2}}; y_1, y_3 \right)} , 
    \end{split}
\end{equation}
which in the $q=y_3$ Schur limit setting simplify respectively to: 
\begin{equation}
    \begin{split}
        &\frac{1}{m_1 !} y_1 ^{m_1 N} Z_{m_1} (y_1 ^{-1}, y_1 y_2) ~, \\
        &\frac{1}{m_2 !} y_2 ^{m_2 N} Z_{m_2} (y_2 ^{-1}, y_1 y_2 ) ~.
    \end{split}
\end{equation}
We also have to consider the hypermultiplet contribution, namely the last line of \eqref{eq: 1/16 Gm m1 m2}:
\begin{equation}
    \prod_{a=1} ^{m_1} \prod_{b=m_1 +1} ^{m_1 + m_2} \fft{\left(\frac{z_a}{z_b} py_2 ^{-1} ; y_3  \right)_\infty \left(\frac{z_a}{z_b} q y_2 ^{-1} ; y_3  \right)_\infty }{\left(\frac{z_a}{z_b} y_2 ^{-1} ; y_3  \right)_\infty\left( \frac{z_a}{z_b} y_1 y_3  ; y_3  \right)_\infty} \fft{\left(\frac{z_b}{z_a} py_1 ^{-1}; y_3  \right)_\infty\left( \frac{z_b}{z_a} q y_1 ^{-1} ; y_3  \right)_\infty }{\left(\frac{z_b}{z_a} y_1 ^{-1} ; y_3  \right)_\infty\left(\frac{z_b}{z_a} y_2 y_3  ; y_3  \right)_\infty }.
\end{equation}
Using $q=y_3$ and $p=y_1 y_2$, this simplifies to:
\begin{equation}
    \begin{split}
        \prod_{a=1} ^{m_1} \prod_{b=m_1 +1} ^{m_1 + m_2} \frac{(1-\tfrac{z_a}{z_b} y_1)(1-\tfrac{z_b}{z_a} y_2)}{(1-\tfrac{z_a}{z_b}y_2 ^{-1})(1-\tfrac{z_b}{z_a}y_1 ^{-1})} = \prod_{a=1} ^{m_1} \prod_{b=m_1 +1} ^{m_1 + m_2} y_1 y_2 = (y_1 y_2)^{m_1 m_2} ~. 
    \end{split}
\end{equation}
After adding up the $m_1 !$ and $m_2 !$ identical contributions to the $G_N ^{(m)}$ expression, we obtain that the Schur limit of \eqref{eq: 1/16 Gm m1 m2} with $q=y_3$ and poles \eqref{eq: 1/16 wz m1 m2 poles} is:
\begin{equation}
G_N ^{(m)}(p,q,\vec y\,)\Big \vert_{w_a = y_1 z_a, w_b = y_2 z_b, q=y_3} = (y_1 ^{m_1} y_2 ^{m_2})^N Z_{m_1}^{(\mathrm{ci})} (y_1 ^{-1} , y_1 y_2) (y_1 y_2)^{m_1 m_2} Z_{m_2 }^{(\mathrm{ci})}  (y_2 ^{-1}, y_1 y_2) ~, 
\end{equation}
exactly matching the Schur index counterpart, \eqref{eq:murthytoafim}. 

One might wonder what happens to the more general $\hat{Z}_{(m_1, m_2, m_3)}(p,q;\vec y\,)$ giant graviton index for $m_1 > 0$ in the Schur limit. While we have not extracted these from Murthy's $G_N^{(m)}$ expansion, primarily due to the presence of the constraint $pq = y_1 y_2 y_3$, it is easy to see that the $m_1>0$ case nevertheless vanishes in the Schur limit by the same mechanism as the vanishing of $\hat{Z}_{(m,0,0)}(p,q;\vec y\,)$.  In particular, when $m_1>0$, the expression for $\hat{Z}_{(m_1, m_2, m_3)}(p,q;\vec y\,)$ contains the prefactor $\Gamma(p; y_2, y_3)$ that vanishes trivially. This further demonstrates how the Schur index can be recovered from the 1/16 BPS index in the giant graviton expansion.

\section*{Acknowledgements}
We wish to thank A.~Arabi Ardehali and S.~Murthy for useful discussions.  This work was supported in part by the U.S. Department of Energy under grant DE-SC0007859.

\bibliographystyle{JHEP}
\bibliography{references}

\end{document}